\documentclass{aa}  

\usepackage{graphicx}
\usepackage{txfonts}
\bibpunct{(}{)}{;}{a}{}{,} % to follow the A&A style
\begin{document}

   \title{H$\alpha$  Doppler shifts in a tornado in the solar corona}

 %  \subtitle{}

   \author{
          %\inst{1}
         B. Schmieder \inst{1}
          \and
          P. Mein\inst{1}
          \and
          N. Mein\inst{2}
           \and
           P. J. Levens \inst{3}
            \and
          N. Labrosse \inst{3}
         \and
          L. Ofman \inst{4}
          %\inst{1}
          %\inst{1}\fnmsep\thanks{Just to show the usage
          %of the elements in the author field}
          }

   \institute{LESIA, Observatoire de Paris, PSL Research University, CNRS, Sorbonne Universit\'es, UPMC Univ. Paris 06, Univ. Paris Diderot, Sorbonne Paris Cit\'e, 5 place Jules Janssen, F-92195 Meudon, France
   \email{brigitte.schmieder@obspm.fr}
         \and
         Observatoire de Paris, 61 avenue de l'Observatoire, 75014 Paris, France
                \and
                SUPA School of Physics and Astronomy, University of Glasgow,
              Glasgow, G12 8QQ, UK
%              \email{p.levens.1@research.gla.ac.uk
              \and
                CUA and NASA Goddard Space Flight Center, Code 671, Greenbelt, MD 20771, USA
              %Goddard Space Flight Centre, 8800 Greenbelt Rd., Greenbelt, MD 20771, USA
              }
         %    University of Alexandria, Department of Geography, ...\\
         %    \email{c.ptolemy@hipparch.uheaven.space}
         %    \thanks{The university of heaven temporarily does not
         %            accept e-mails}
       %  CISL/HAO, National Center for Atmospheric Research, P.O. Box 3000,Boulder, CO 80307-3000, USA

   \date{Received ...; accepted ...}

% \abstract{}{}{}{}{} 
% 5 {} token are mandatory
 
  \abstract
   {High resolution movies in 193 \AA\ from the Atmospheric Imaging Assembly (AIA) on the \textit{Solar Dynamic Observatory} (\textit{SDO}) show  apparent rotation in the leg of a  prominence observed during a coordinated campaign. Such structures are commonly referred to as  tornadoes. Time-distance intensity diagrams of the AIA  data show the existence of oscillations  suggesting that the structure is rotating.  }
   {
   %AIA observations in the 193 \AA\ passband show the prominence as dark structure due to absorption of the background coronal emission. 
   The aim of this paper is to understand if the cool plasma  at chromospheric temperatures inside the tornado is rotating around its central axis.} 
   %or if the motions seen in AIA movies are caused by oscillations.}
   {The  tornado was  also observed   in  H$\alpha$ with a cadence of 30 seconds  by  the MSDP spectrograph, operating at the Solar Tower in Meudon. The MSDP provides sequences of   simultaneous spectra in a 2D field of view from which    a cube of Doppler velocity maps  is  retrieved. 
    } 
   {
  The  H$\alpha$  Doppler maps show  a  pattern with alternatively blueshifted and redshifted  areas of 5 to 10\arcsec\ wide.  Over time  the blueshifted  areas become redshifted and vice versa,  with a quasi-periodicity of 40 to 60 minutes. Weaker amplitude oscillations with periods of 4 to 6 minutes are superimposed onto these large period oscillations.     }
 {   The Doppler  pattern observed in H$\alpha$  cannot be  interpreted as  rotation of  the cool plasma inside the tornado.  The  H$\alpha$ velocity observations give strong constraints on the possible interpretations 
   of the AIA tornado.
  }


   \maketitle
%
%________________________________________________________________

\section{Introduction}

%__________________________________________________________________

%With the new armada of spacecraft observing the Sun (Hinode, SDO) we  are able to have an accurate view of the solar phenomena occurring in the corona. 
The term solar tornadoes has been used to describe apparently rotating magnetic structures above the solar limb, as seen in high resolution images and movies from the  \textit{Atmospheric Imaging Assembly} (AIA)  aboard the \textit{Solar Dynamics Observatory}  (SDO) \citep{Lemen2012}.
%\citep[AIA, \textit{SDO};][]{Lemen2012}.
%Solar tornadoes have been the topic of a number of publications recently, since the launch of the \textit{Solar Dynamics Observatory} (\textit{SDO}) spacecraft with its high resolution imager, the Atmospheric Imaging Assembly (AIA) .
Movies obtained with high spatial and temporal resolution have enabled the discovery of the incredible dynamic nature of prominences.  Even in quiescent prominences, apparently rotating, tornado-like structures, apparent upflows and downflows in quasi-vertical structures, and rising bubbles  are seen \citep{Dudik2012,Orozco2012,Wedemeyer2013,Berger2014,Su2014}.  
Spectroscopy is  necessary to analyse the real plasma motion and physical conditions of the tornado.   An example of tornado  rotation around its axis was detected in hot plasma  (T $>$  $10^6$ K) surrounding the prominence legs by 
the Extreme-ultraviolet Imaging Spectrometer (EIS) on \textit{Hinode} \citep{Kosugi2007} using Doppler shifts in a number of coronal lines \citep{Su2014,Levens2015}. 
We ask here whether or not the cool material (T  $\sim 10^4$ K) inside the tornado, which is visible in chromospheric lines, also presents signatures of rotation and, if so, with what velocity?
%   {\bf We would like to know if the cool matter  (10$^4$ K) inside  tornado visible in chromospheric lines is also turning as well?}

\citet{Wedemeyer2013} showed some examples of tornado-like structures observed with the Swedish Solar Telescope (SST) in the H$\alpha$ line and posed the question: Are they the legs of prominences? 
Their first example is an observation of a filament on the disc. Blueshifts and redshifts are observed along the axis of the structure that could give the impression of  a rolling flux rope. Their other example is  a prominence observed above the limb. Torsional motions, such as those observed  during an eruption, are detected along fine threads. These two examples were not conclusive concerning these apparent tornado-like structures. New ground-based observations were required to investigate these prominence motions further.\\
%All what is protruding the solar limb is not obviously a prominence  with its characteristics.   Sometimes it is difficult to identify a  prominence  as a filament  overlaying a magnetic inversion line, a few days before its passage over the limb. 
\citet{Poedts2015} explored the oscillations seen in tornadoes observed with coronal AIA filters (171 \AA, 211 \AA, 193 \AA). They concluded that they could see two kinds of patterns depending on the time.  During the first 
phase the tornadoes were rising with a twisting pattern,  while  later the  tornadoes stopped their rise and their  oscillations were interpreted as due to Magneto-Hydro-Dynamic (MHD) kink waves with one end fixed in the photosphere and an open end at the top, or by the rotation of two tornado structures during a quasi-periodic phase.\\
 \citet{Panasenco2014} explained the apparent vortical motion in prominence spines and barbs exhibited in \textit{TRACE} and \textit{SDO}/AIA 171 \AA\ images and movies as counterstreaming flows  giving the illusion of rotation. The apparent rotational motion is only observed in 2D projection at the limb in the plane of the sky. 
 The authors claimed, "the constant counterstreaming motion of the prominence plasma along the thin threads, especially when they connect the vertical parts of the prominence (legs and barbs) to the much more horizontal spine, creates an effect that the eye associates with rotation". %"The counter streaming motion of prominence along threads, in fact along bushes of threads, especially when they connect the vertical parts of the prominence (legs) to the horizontal spine can create the effect of rotation". 
 They concluded that the tornado-like structure  oscillates, but does not rotate.

 {Oscillations in prominences are  frequently observed and are important for prominence seismology because of the possible role of MHD waves in heating the prominence material \citep{Ofman1998,Ofman2015}. With the high temporal and spatial resolution of recent solar telescopes, it has been possible to resolve small-scale oscillations and waves \citep{Engvold2008}.  Transverse oscillations have been reported  recently using \textit{Hinode}/SOT and  \textit{SDO}/AIA \citep[see papers of][]{Okamoto2007,Okamoto2015,Schmieder2013}.  These oscillations,  which are  commonly observed as  density fluctuations, concern either horizontal fine threads, with periods of the order of 10 to 15 minutes \citep{Okamoto2015}, or feet of prominences, with periods of 5 minutes
 \citep{Schmieder2014}. \citet{Ofman2015} interpret the   transverse oscillations  observed in a feet of prominence   as due to non-linear fast magnetosonic waves. They also explore non-linear gravitational MHD oscillations of heavy material in prominence feet, supported by a dipped magnetic field structure using a 2.5D MHD model and find lesser agreement with these global waves owing to their longer periods. 
 The transverse oscillations  in fine horizontal threads have been interpreted as being Alfv\'{e}n waves, coupled with kink waves \citep{Antolin2015}. However, these oscillations and waves  have  smaller periods as the  typical tornado period of 1.5 to 2 hours.}
 
 An interesting spectroscopic approach to the problem of rotation was taken by \citet{Orozco2012} looking at  a tornado for one hour with the Tenerife Infrared Polarimeter (TIP) instrument operating at the Vacuum Tower Telescope (VTT) in the Canary Islands.
 They used the \ion{He}{I} %10830 \AA\, triplet
 infrared multiplet. Surprisingly the two components  at 1082.909 nm and 1083.029 nm have an  opposite behaviour in terms of Doppler shifts. Both components have a low intensity signal, a weak  line width, and are optically thin.
  These authors concluded that there are possible rotational signatures at the edges of the tornado because in the Doppler shift versus time diagrams the cells of blueshifts are consistently located at the  right edge of the leg, and the redshifts at the left edge of the leg. However this interpretation needs to be supported by more observations with similar characteristic Doppler shifts.

{The apparent upflows and downflows in prominence legs do not prove that the structure  is vertical.} Using H$\alpha$ spectra obtained with the Multi-channel Subtractive Double Pass (MSDP) instrument in the Meudon Solar Tower 
it was shown that in prominences the velocity vectors were not aligned with  the apparent vertical  intensity structures, as the movies suggest, but have a significant 
angle with  respect to the vertical. This  suggests the existence of a more or less  horizontal  magnetic support rather than vertical flows \citep{Schmieder2010}.  In an analysis of a hedgerow prominence, \citet{Chae2010} suggested a similar magnetic support. The 
  descending observed  knots are basically supported  against gravity by horizontal magnetic fields  even when they descend, and the complex variations of their descending speeds should be attributed to small imbalances between gravity and force of magnetic tension.

% but has  not yet  been confirmed in cooler plasma  (\cite[Levens et al. 2015]{Levens2015}).
%The IRIS spectrograph \citep{DePontieu2014} provides tremendous data on prominences, quiescent to eruptive prominence observed in the chromospheric Mg II lines  \citep{Schmieder2014,Liu2015}.
%The highly  dynamic   plasma observed in quiescent prominences  could  answer  the question of short scale height of the plasma pressure compared with the common height of prominences \citep{Schmieder2014}. 

%All the theoretical models are based on static structures  \citep{Aulanier1998,vanBallegooijen2004,Dudik2008,Mackay2010}. The plasma  would be  sustained in a  pile  of dips in sheared arcades or untwisted flux ropes due to magnetic tension force or the presence of tangled magnetic field on small scales \citep{Lopez2006,Lites2010,vanBallegooijen2010}. This is consistent with previous measurements of the magnetic field in prominences. Polarimetry of prominences was  achieved in the 1980s \citep{Leroy1984} which  showed that the 
%prominence magnetic field vector was almost horizontal (60 to 90 degrees from the vertical). The inversion codes developed during this time period involved only the Hanle effect, neglecting the Zeeman effect \citep{Bommier1994}. It appears that the second effect cannot be neglected. New inversion codes have been  developed to  include all those effects and are now being applied to new sets of full Stokes  vector observations \citep{Casini2003,Lopez2007,Lites2014}.  
The French telescope {\it T\'elescope H\'eliographique pour l'Etude du Magn\'etisme et des Instabilit\'es Solaires} (THEMIS) in the Canary Islands with its MulTi-Raies (MTR) mode has been observing prominences during international campaigns since 2012. More than 200 prominences have been observed in the \ion{He}{I} D$_3$ line. Statistical work has been presented \citep{Lopez2015} and  case studies have been published \citep{Schmieder2013,Schmieder2014}.  The main result is that the magnetic field is mainly horizontal in prominences.  Recently the magnetic field of  tornado-like structures observed by THEMIS has been analysed  in 2D maps \citep{Schmieder2015,Levens2016a,Levens2016b}.  Their histogram shows a primary horizontal component with two secondary peaks that are not easy to interpret. The field strength could reach 40 gauss in some parts of a tornado.
 From the theoretical point of view we can quote the MHD model of \citet{Luna2015}, which was based on the modelling of a vertical cylinder with a vertical field along its central axis and a more and more helical field as closer to the periphery. It is difficult to model such a structure and obtain the corresponding Stokes parameters that have been observed.  

However, as has been mentioned, tornadoes are often  considered to be legs of prominences \citep{Wedemeyer2013,Levens2016a}. Many static models proposed that legs or barbs of prominences  are piles of dips in magnetic field lines supporting the cool plasma \citep{Aulanier1998,Dudik2008,Heinzel1999,Mackay2010}. More recently  simulations of filament  formations  by condensation have  supported this idea \citep{Xia2014,Terradas2015}. 

In this paper we present   a tornado-like prominence observed by AIA  in 193 \AA\ on September 24, 2013, where it looks like it is rotating as a vertical structure with a characteristic period of  around 90 minutes  (Section 2 and movie), and  in  H$\alpha$ with observations made by the MSDP spectrograph (Section 3).  Our aim is  to compute the velocities along the line of sight of  the cool plasma to see if we observe similar rotational signatures to those seen with AIA. In the last section, we discuss possible interpretations of the observed oscillatory behaviour of the H$\alpha$ Doppler shifts for the AIA   tornado. 

%Figure 1
  \begin{figure}
   \centering
   \includegraphics[width=\hsize, clip=true, trim=2cm 2cm 2cm 2cm]{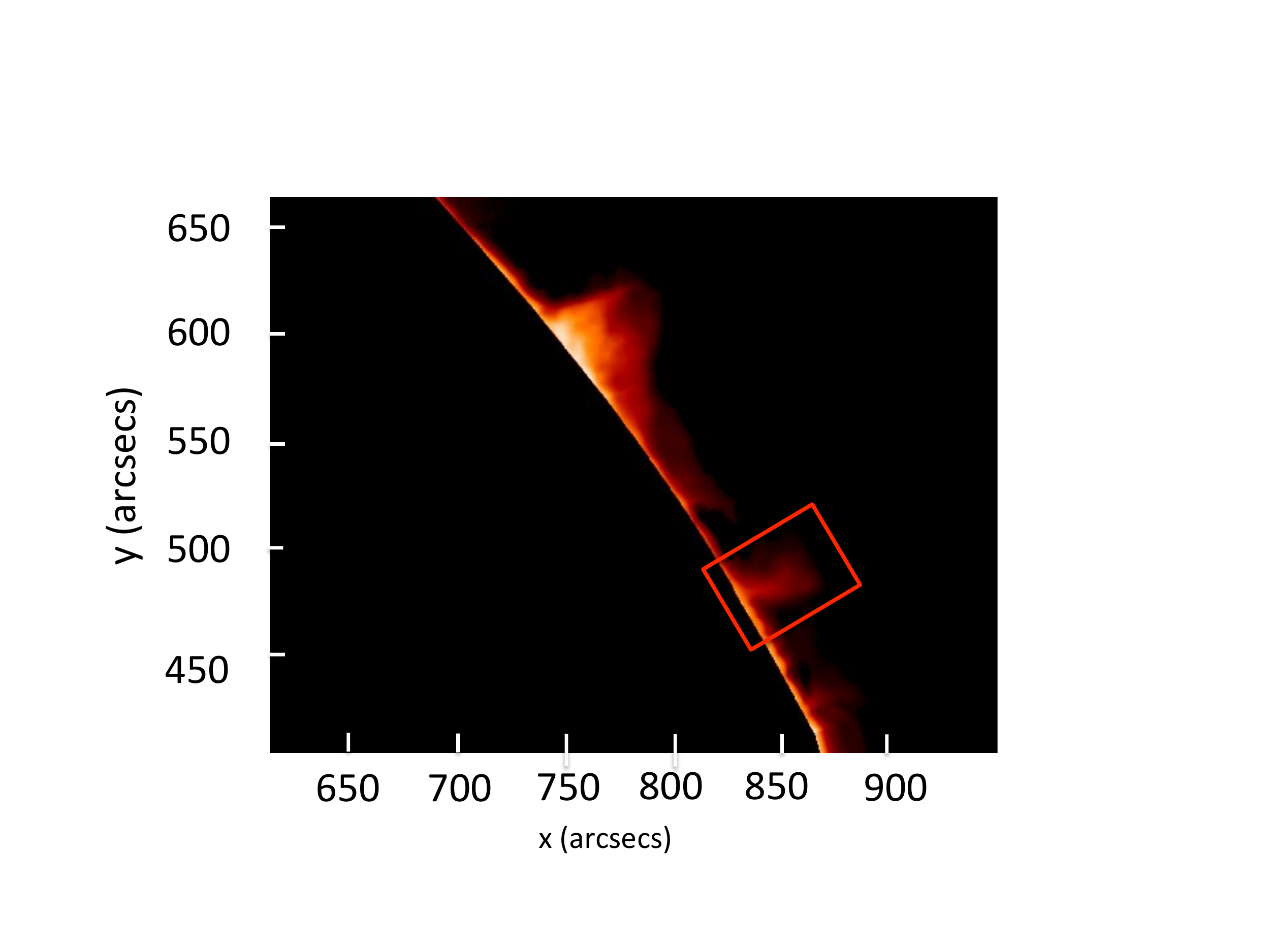}
    \includegraphics[width=\hsize, clip=true, trim=2cm 2cm 2cm 2cm]{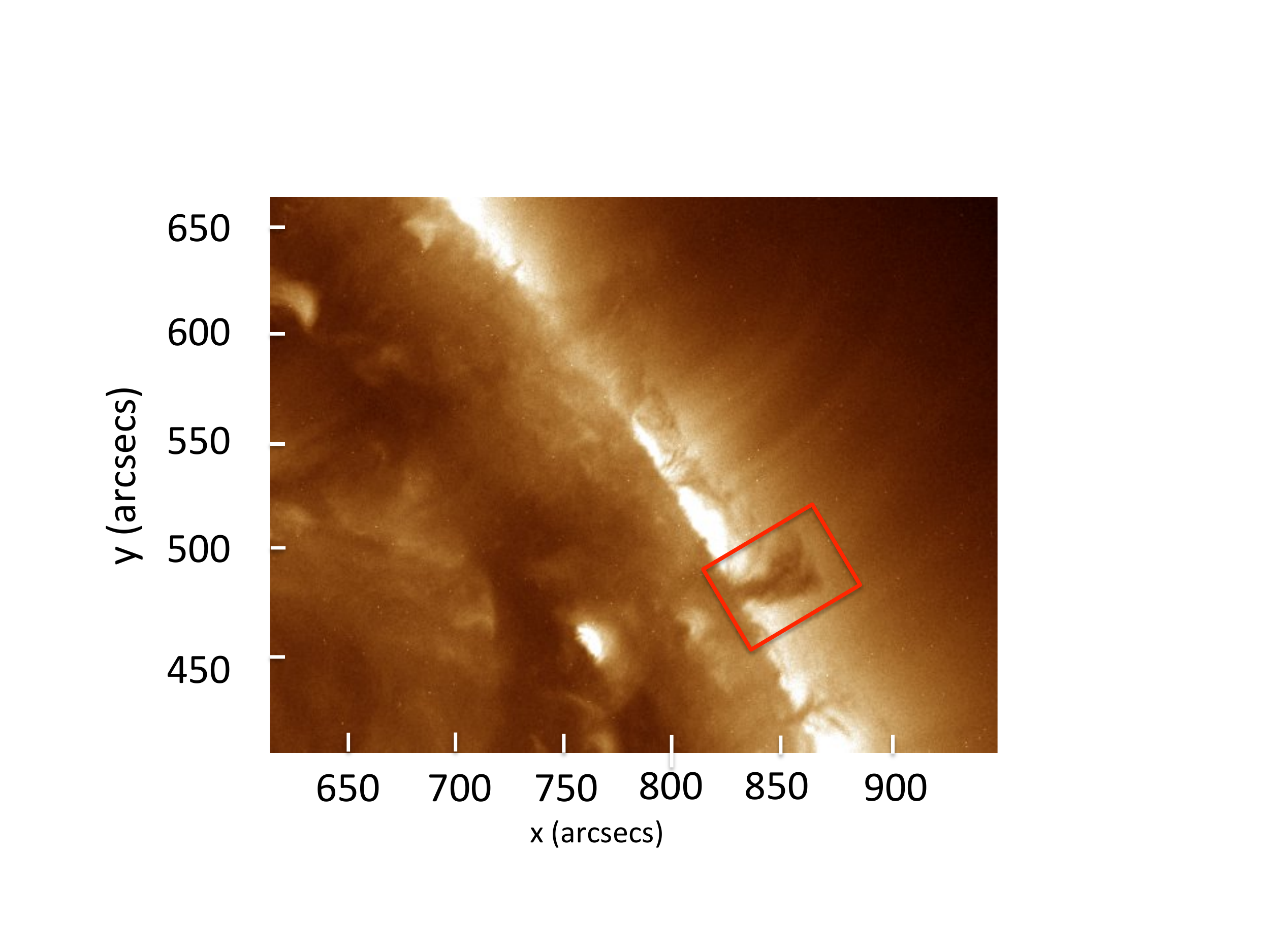}
 \includegraphics[width=\hsize, clip=true, trim=2cm 2cm 2cm 2cm]{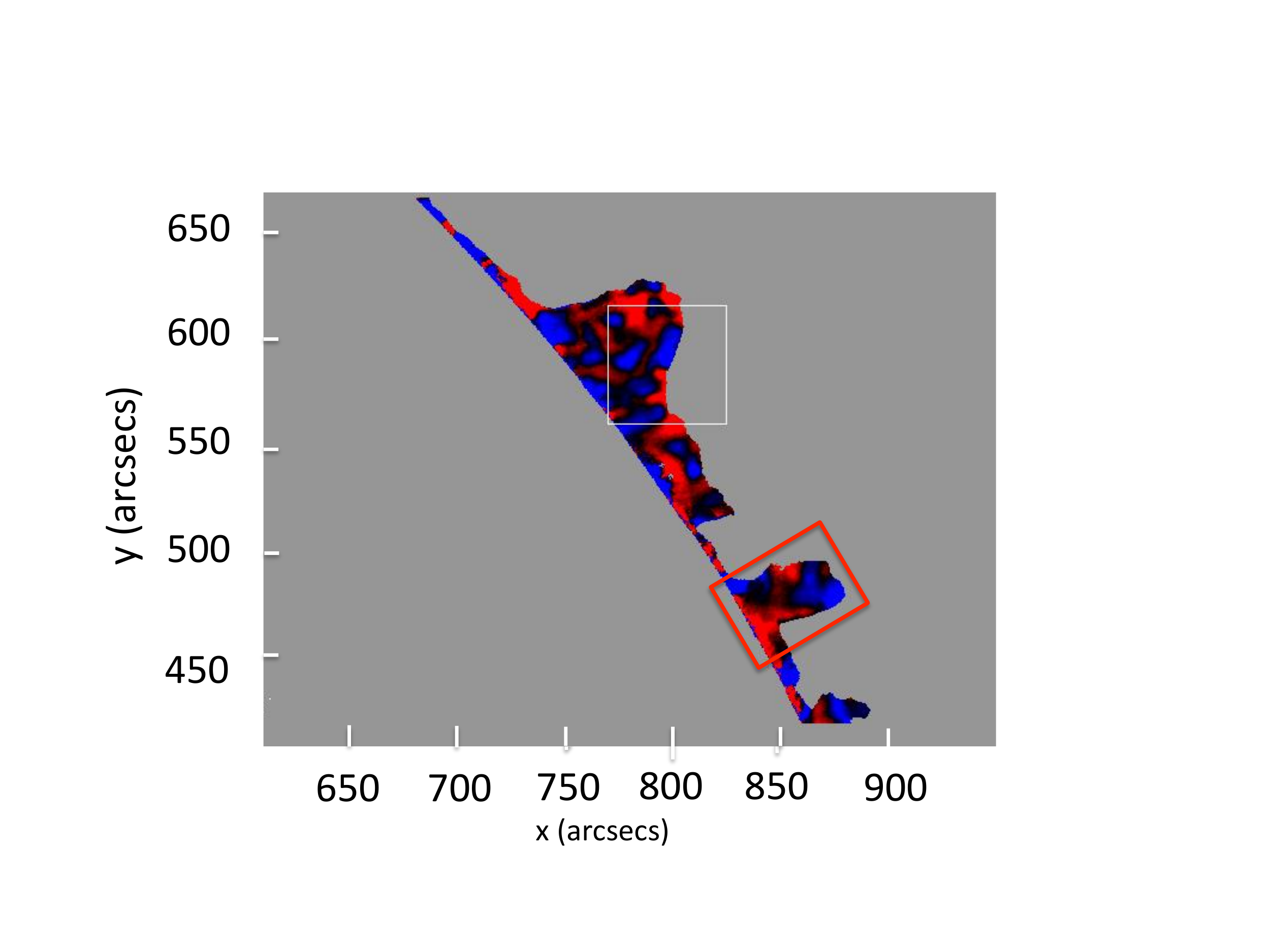}
 \caption{Prominence observed on September 24, 2013  at 12:22 UT.  \textit{Top panel:} H$\alpha$ image from the MSDP instrument operating on  the solar tower in Meudon.  \textit{Middle  panel:} 
 \textit{SDO}/AIA 193  \AA\ image. The prominence   in 193 \AA\, is observed in absorption.  \textit{Bottom panel:}     H$\alpha$ Doppler shift map. The  tornado is in the  red box, which corresponds to the field of view of the top panels of Figures \ref{profil}, \ref{Doppler1}, \ref{Doppler2}, and \ref{Doppler3}.  The white box  indicates the  part of  the  prominence  observed by IRIS \citep{Schmieder2014}. }
         \label{MSDP}
   \end{figure}
%

 % \begin{figure}
  % \centering
   % \includegraphics[width=\hsize]{image_time_distance_2}
 %  \includegraphics[width=\hsize]{tornado_23sept_2013_Ha}
   % \caption{Prominence observed on September 23, 2013   at  08:20 UT  with  SDO/AIA  193 \AA,. The dashed lines show where the cuts of Figure \ref{time} have been achieved.  }
      %   \label{AIA}
  % \end{figure}
%
%Figure 2
 \begin{figure}
   \centering
     \includegraphics[width=0.7\hsize, clip=true, trim=4.4cm 3cm 0 0]{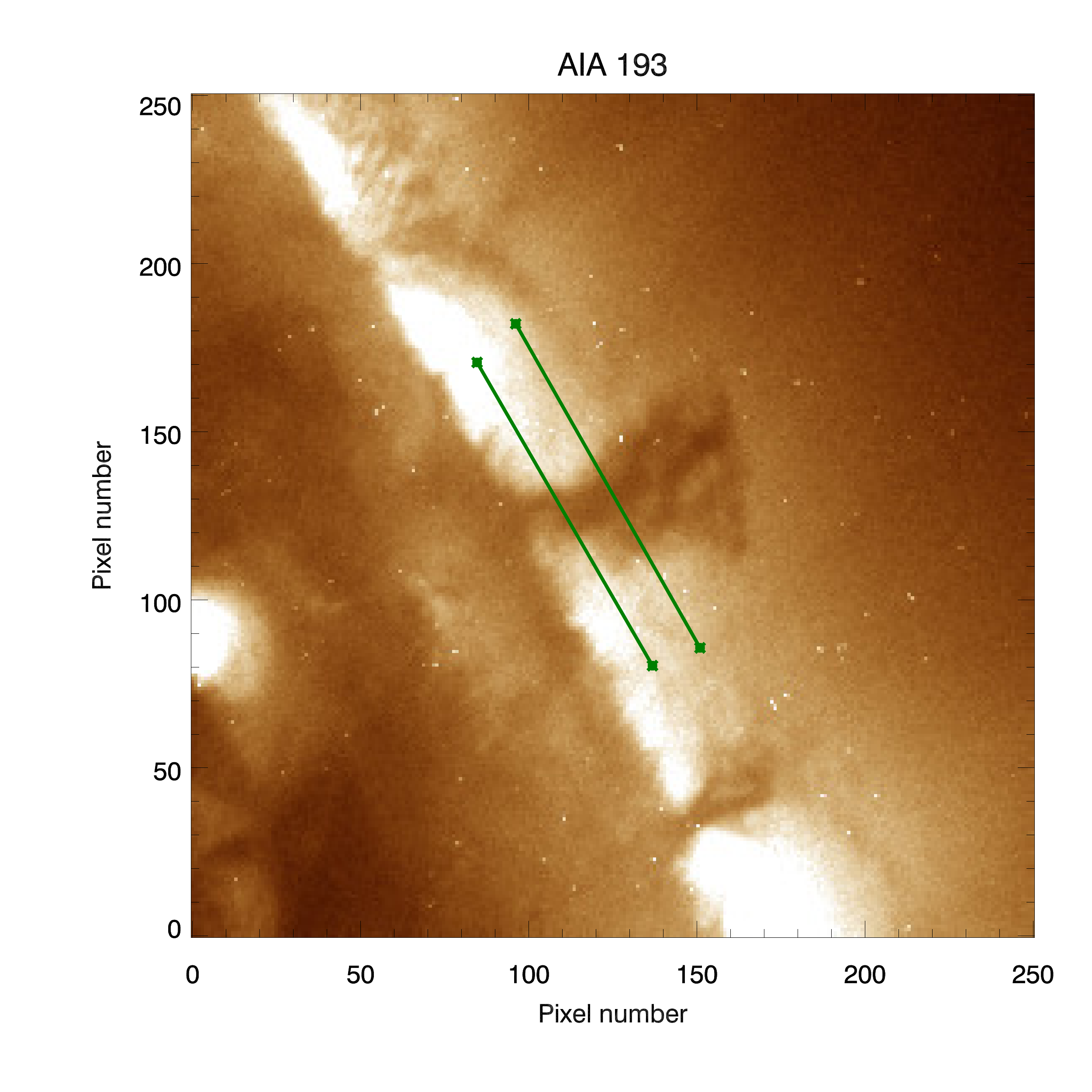}
     \includegraphics[width=\hsize, clip=true, trim=3cm 0 1cm 0]{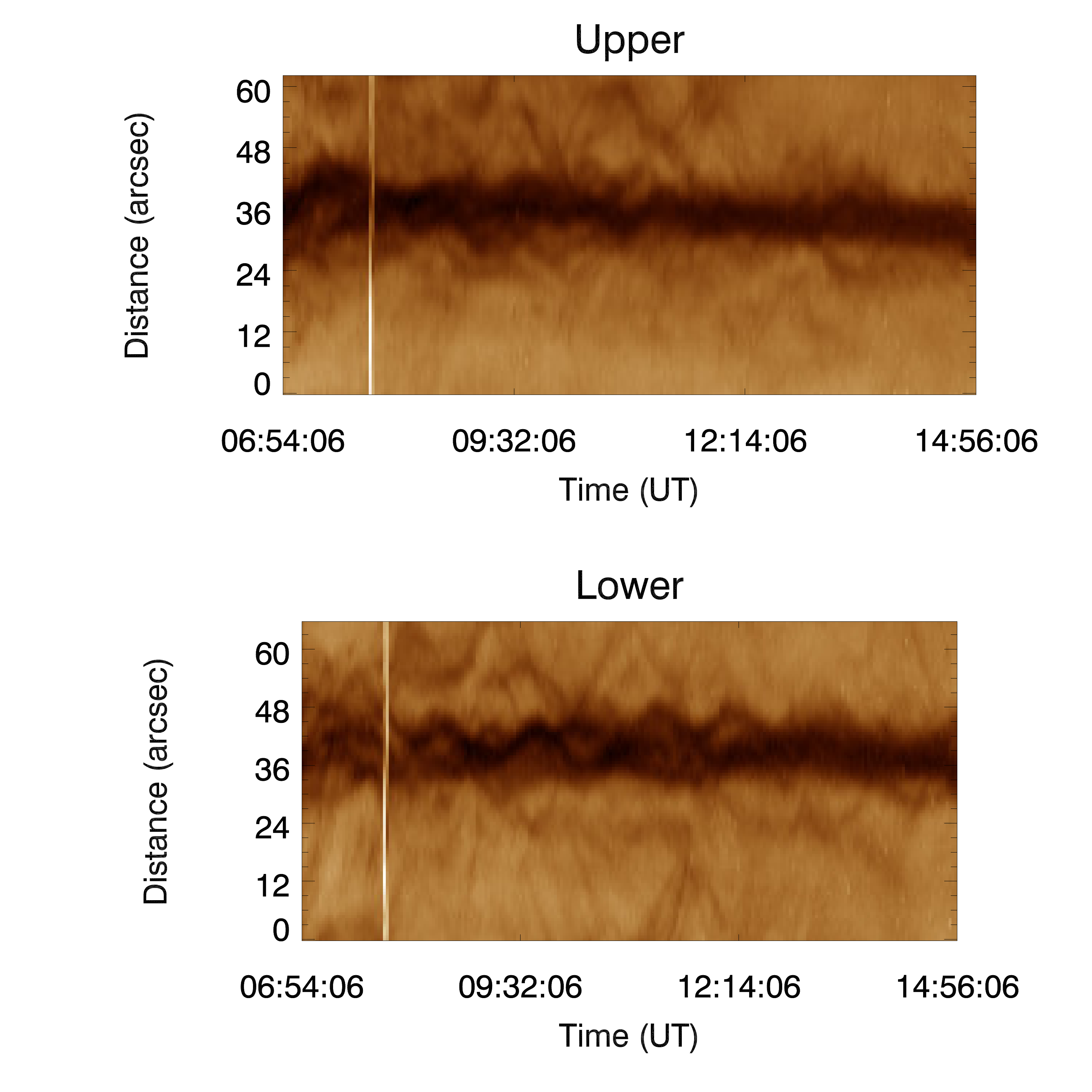}
       \caption{Prominence observed on September 24, 2013   by \textit{SDO}/AIA in the 193 \AA\ window. \textit{Top panel:} Image from AIA at  06:54 UT. Field of view is approximately 200\arcsec\ $\times$ 200\arcsec. The green lines show where the cuts in the bottom panels were measured. \textit{Bottom panels:} Time-distance  intensity diagrams for the two cuts  between 06:54 UT and 14:56  UT (8 hours). The distance between the two cuts is approximately 10 Mm.}
% The time step is 2 minutes. The unit on the $y$ axis is AIA pixel number = 0.6 \arcsec. The distance between the two cuts is approximatively 10 Mm.}
         \label{time}
   \end{figure}
%
 
% Figure 3
  \begin{figure}
   \centering
   \includegraphics[height=0.9\hsize,clip=true,trim=4cm 3cm 2cm 5cm,angle=-90]{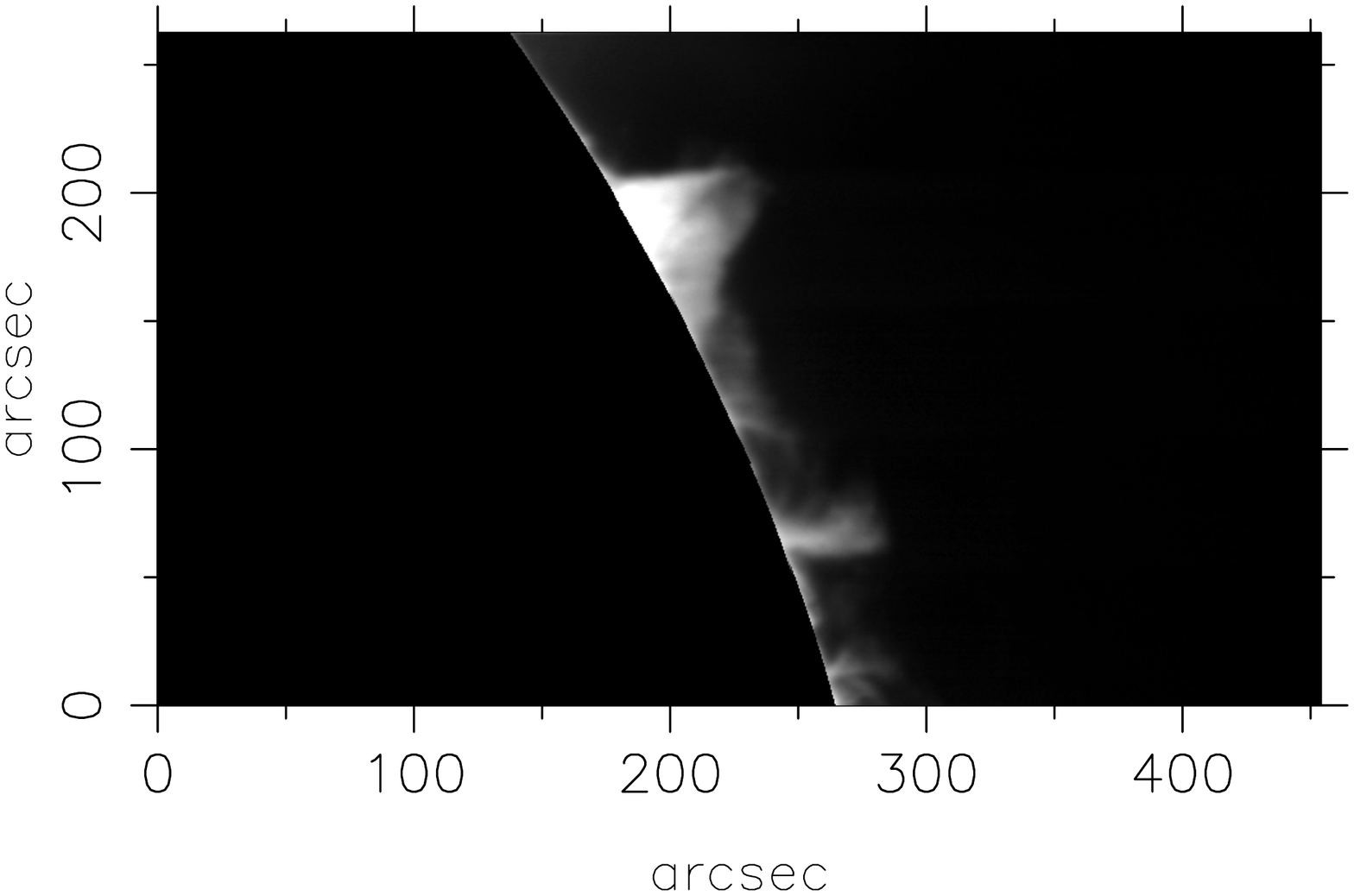}
  \includegraphics[height=0.9\hsize,clip=true,trim=4cm 2cm 2cm 4cm,angle=-90]{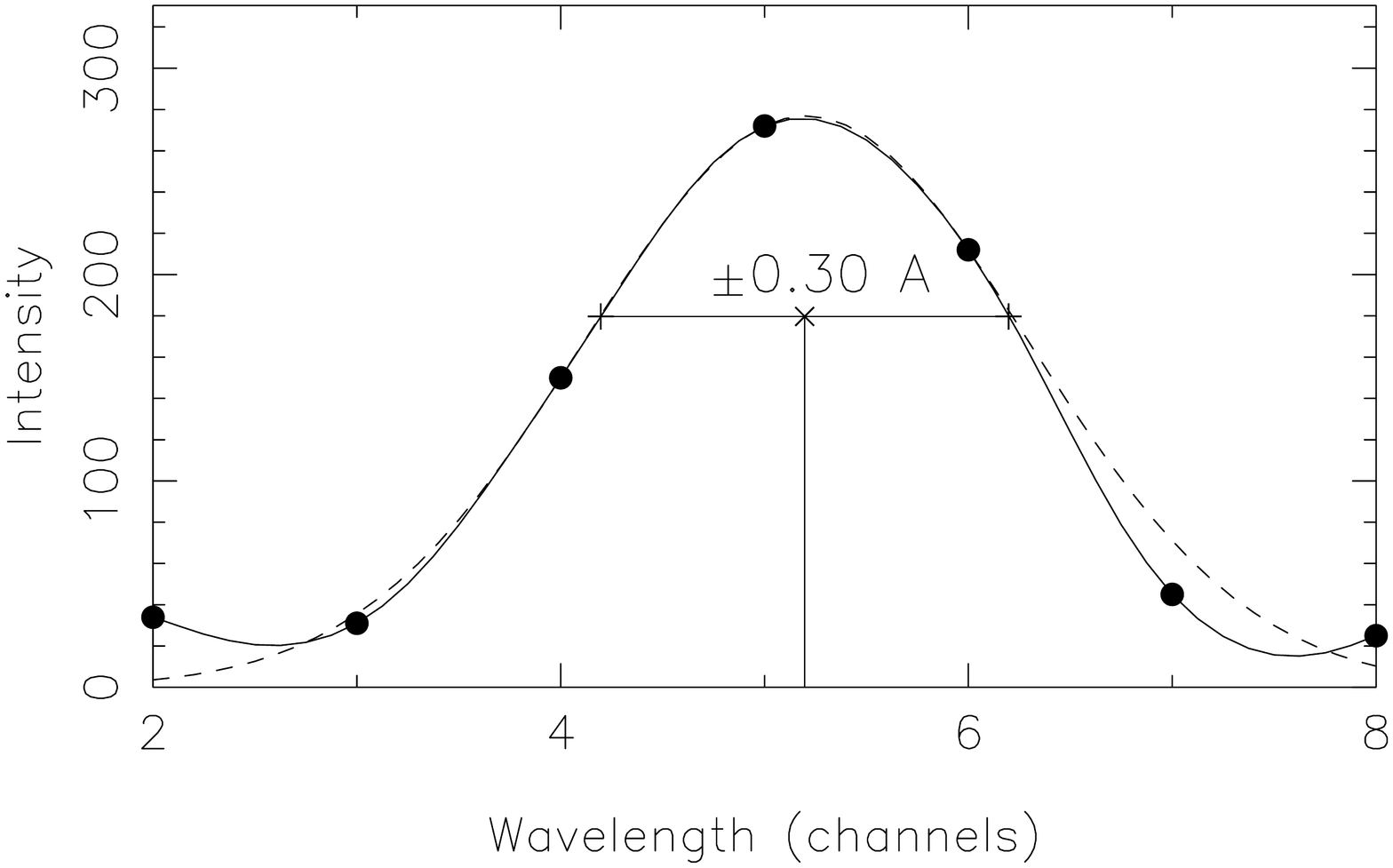}
      \caption{\textit{Top panel:} Full field-of-view image  (450\arcsec $\times$  260\arcsec) of the MSDP in non-heliographic coordinates (x,y).
      \textit{Bottom panel:}        Example of interpolation of MSDP $H\alpha$ profiles  by the cubic method (solid line) and by a Gaussian  function (dashed line) and 
determination of Doppler shifts via a bisector method.}
\label{bissector}
\end{figure}
   
   %Figure 4
 \begin{figure}
   \centering
    \includegraphics[width=0.8\hsize,clip=true,trim=4.5cm 4.5cm 4.5cm 3cm]{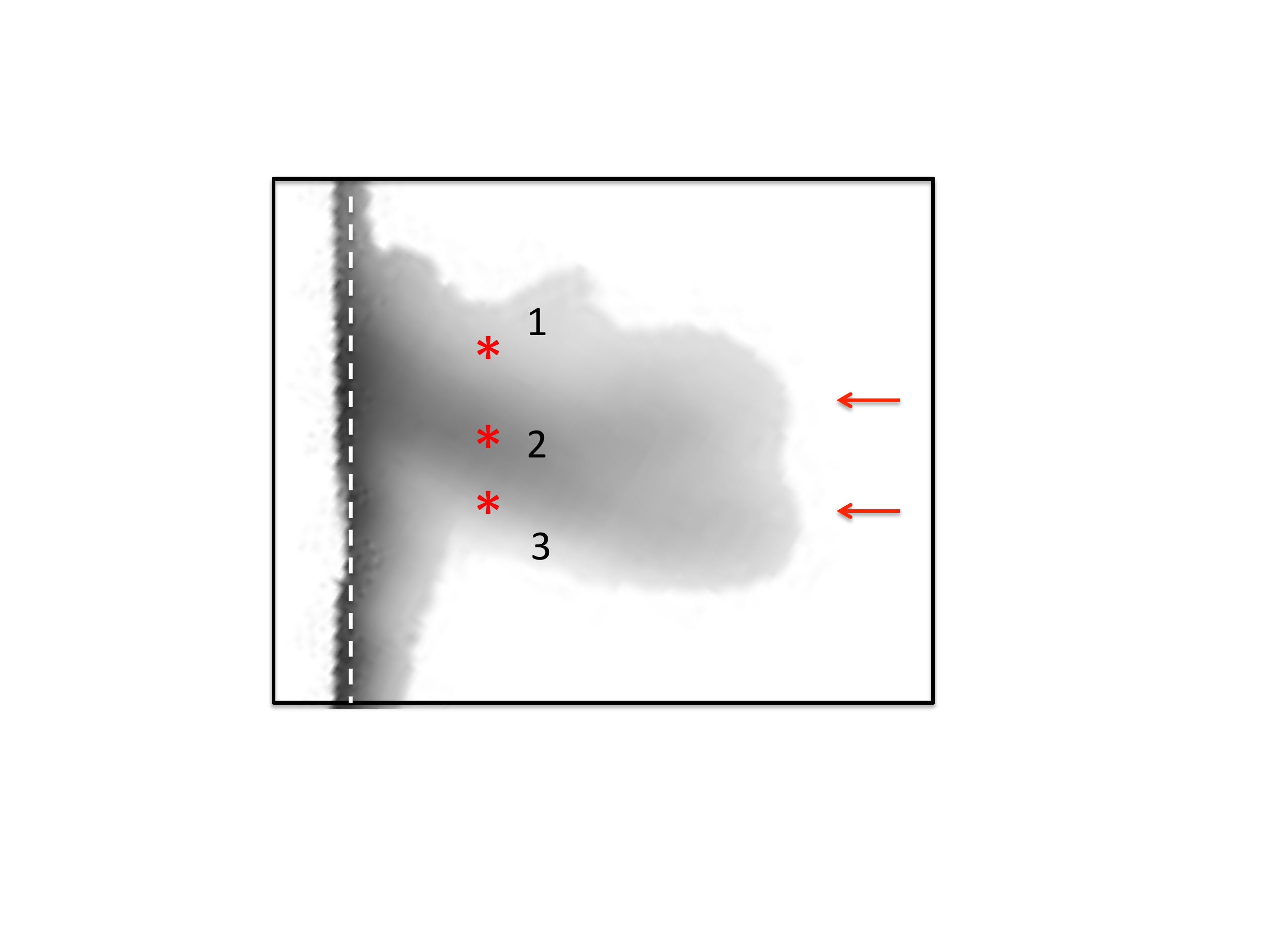}
       \includegraphics[width=0.8\hsize]{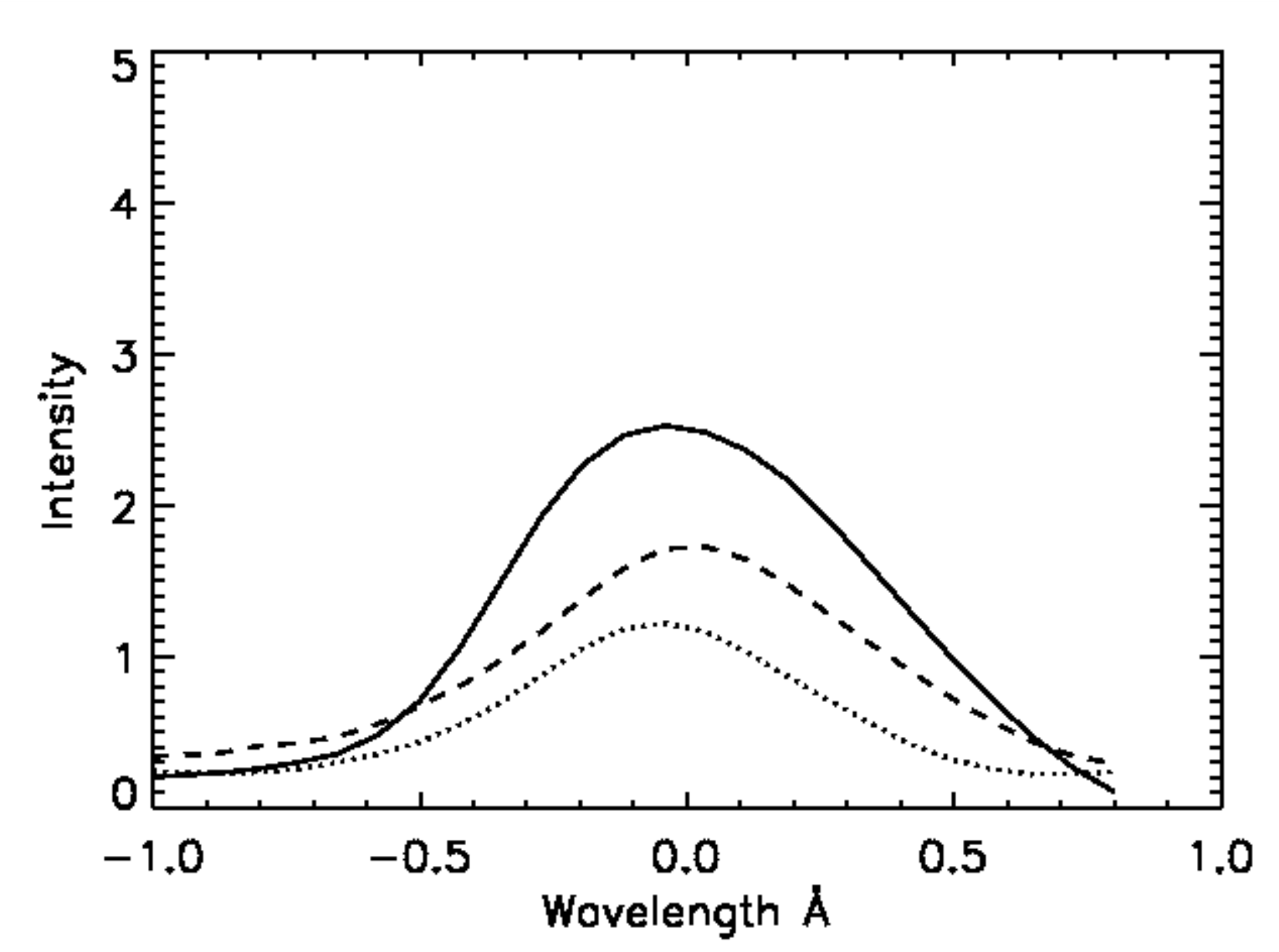}
      \caption{\textit{Top panel:} H$\alpha$  integrated  intensity  map of the tornado observed with the Meudon MSDP.  The field of view is 70\arcsec $\times$ 45\arcsec. The dashed line approximately represents the limb.  The two red arrows point out the double structure of the tornado at its top. 
      Asteriks 1, 2, and 3 are the  selected pixels for showing examples of  profiles. 
\textit{Bottom panel:} H$\alpha$ profiles: pixel 1 dashed line, pixel 2  solid line, and pixel 3 dotted line.
%corresponding to the central part (solid line), south side (dashed line) and north side (dotted line) of the tornado.
The unit of the intensity  is 10$^{-6} $ erg s$^{-1}$ cm$^{-2}$ sr$^{-1}$ Hz$^{-1}$. }
         \label{profil}
   \end{figure}

% plasma build-up in `dips' in the magnetic field \textbf{ETC ETC MORE IDEAS AND REFERENCES PLEASE}.
\section{AIA tornado}
In the 193 \AA\ spectral window of AIA the tornado, or leg of the prominence, is seen as a silhouette in absorption on September 24, 2013. This absorption is mainly from neutral hydrogen and neutral/ionized helium seen against a bright background  \citep{Schmieder2004,Anzer2005} (Figures \ref{MSDP}, \ref{time}).  The tornado height  is about 45 Mm. Its width varies with time and height above the photosphere. Sometimes it splits into two or more threads and shows lateral extension, as seen in Figure \ref{time}.  Generally this tornado is narrower near the footpoint and wider at the top, as is noted by \citet{Poedts2015}. In order to study the temporal dynamics of the tornado, we construct time-distance diagrams at two different heights above the limb. The location of these cuts is shown with green lines in Figure \ref{time}. The first cut is located at a height of 20 Mm and the second cut is around 30 Mm. The bottom panels of Figure \ref{time} show the time-distance diagrams. There are apparent quasi-periodic transverse displacements of the axis during the interval of time (8 hours).
 The period and amplitude of the displacement is difficult to compute because many different structures are visible.% (80-20 = 60 time steps =120 minutes in cut 0). t
 The period starts off around 80 minutes, and then increases  to about 110 minutes at later times (from the lower cut). A mean period of 90 minutes is a reasonable value.
The tornado  looks as if it is a double structure at the top, which is similar to that seen in \citet{Poedts2015}.
%70-30 - 40 x2 minutes x 2 =  160 minutes ,  -60-38=  22 it leads to a period of 90 minutes -- for the other cut  40-10 = 30 soit 120 minutes. The different threads are in phase or opposition of phase.  
According to the amplitude of the oscillations, we estimate a velocity of 7 km s$^{-1}$.

%?igure 5

 \begin{figure*}
   \centering
  \includegraphics[width=0.8\hsize,clip=true,trim=0 1cm 0 0]{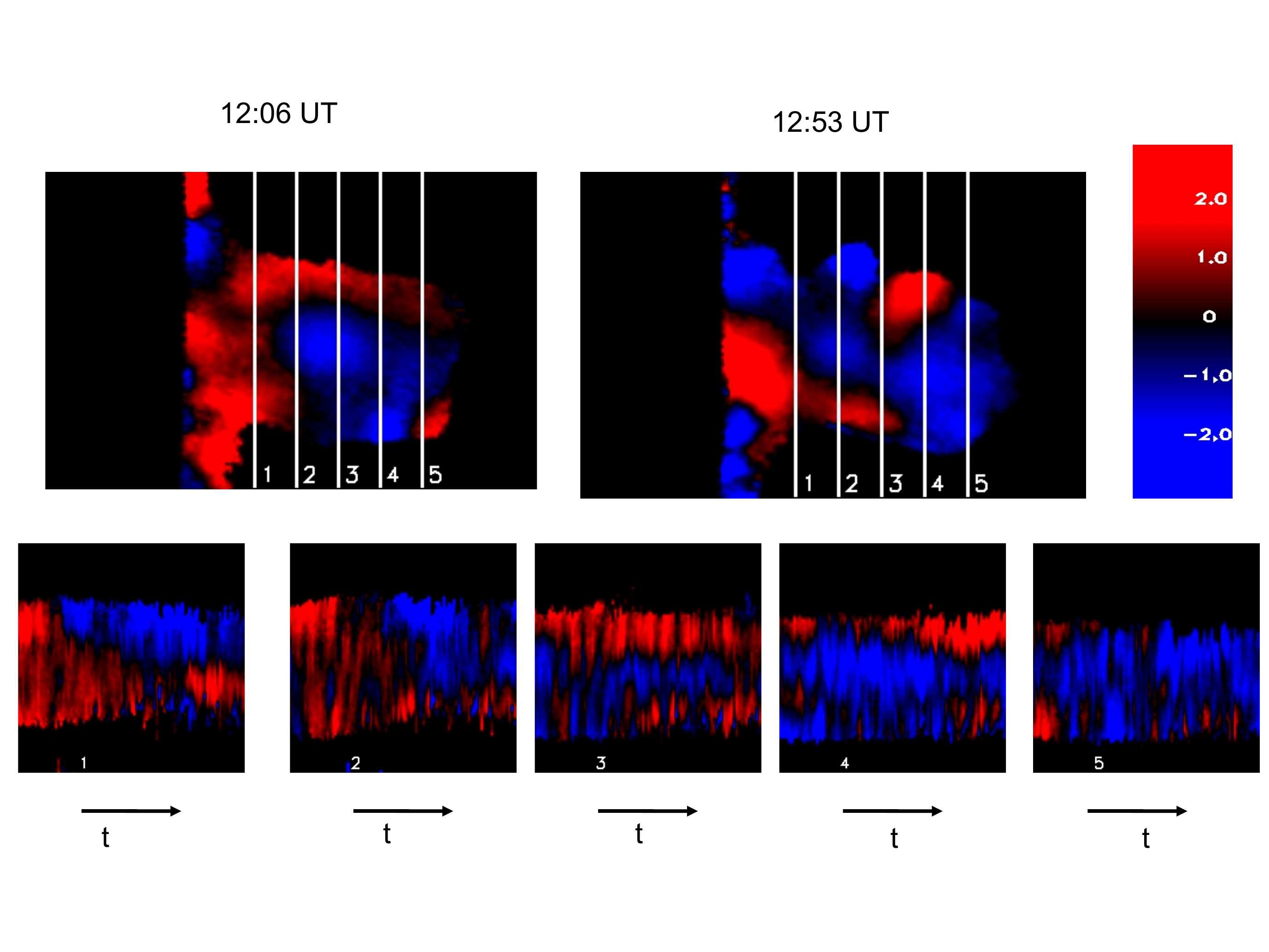}
      \caption{Prominence observed on September 24, 2013 by the MSDP operating on the solar tower in Meudon.% (56 x44 arc sec ), distances entre les coupes 5"figure 5 champ total 70"x45", distance 6")  
 \textit{Top panels:} Doppler shift maps at 12:06 UT and 12:53 UT and the colour bar (the  units are in km s$^{-1}$). Between blue and red, the black colour corresponds to zero velocity. Blue/red are blue and redshifts, \textbf{respectively}. The limb is vertical on the left of the images. The FOV is 70\arcsec\ $\times$ 45\arcsec. \textit{Bottom panels:} Time distance  Doppler shift diagrams over 58 minutes for the five cuts (1 to 5) shown in the upper panels at a distance of 5\arcsec\ apart.  The time step is 30 seconds.  }
         \label{Doppler1}
   \end{figure*}
   
   %Figure 6
   
       \begin{figure*}
   \centering
   \includegraphics[width=0.8\hsize,clip=true,trim=0 1cm 0 0]{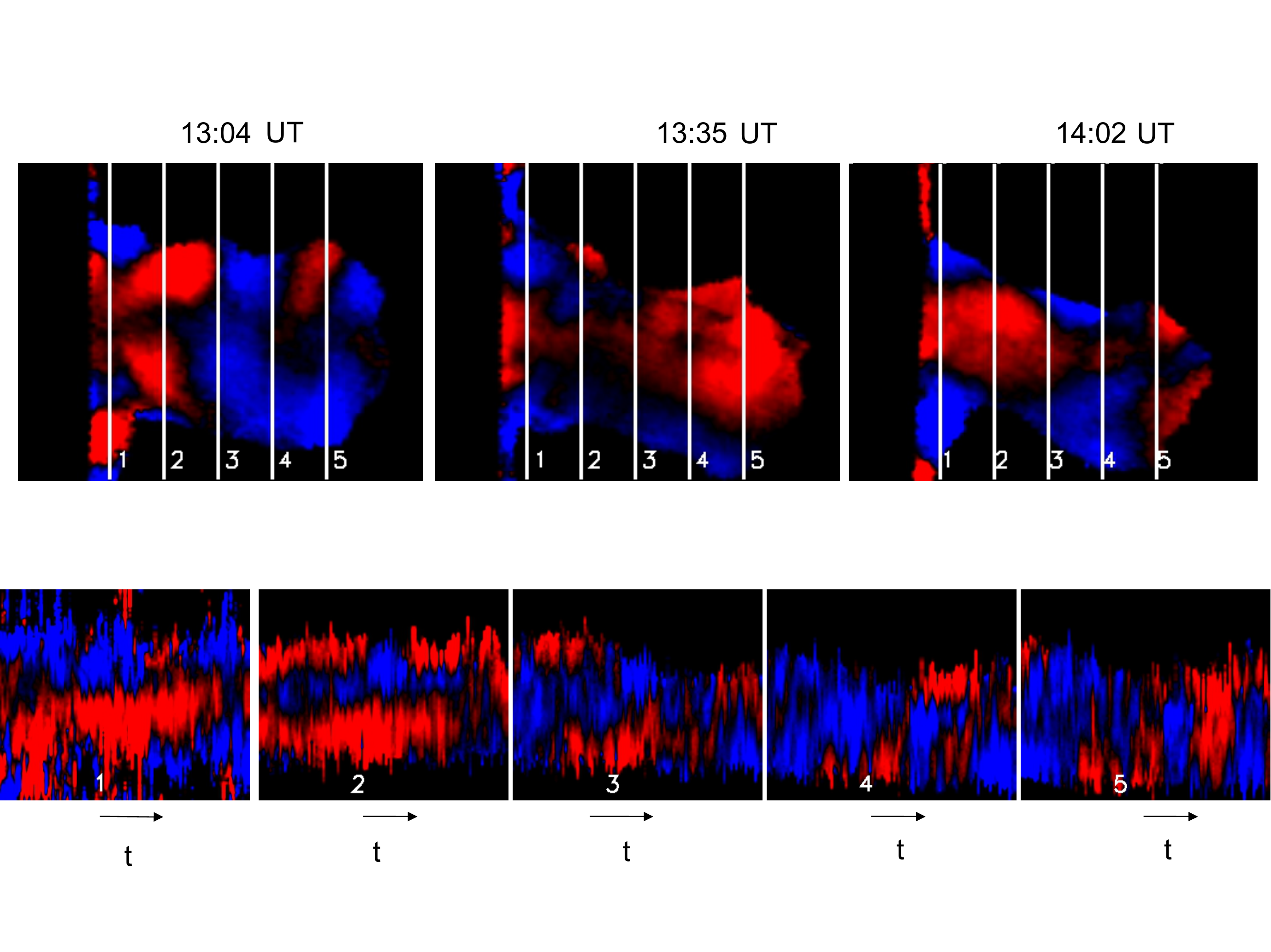}
      \caption{Prominence observed on September 24, 2013 by the MSDP operating on the Solar Tower in Meudon. \textit{Top panels:} Doppler shift maps at 13:04 UT, 13:35 UT and 14:02 UT. The limb is vertical on the left, and the field of view is 70\arcsec $\times$ 45\arcsec. \textit{Bottom panels:} Time distance diagrams over 47 minutes for the five cuts (1 to 5) shown in the upper panels at a distance of 6\arcsec\ apart. The time step is 30 seconds.  See the colour bar in Figure \ref{Doppler1}.}
         \label{Doppler2}
   \end{figure*}
   
 %  Figure 7
   
  \begin{figure}
   \centering
    \includegraphics[width=6cm]{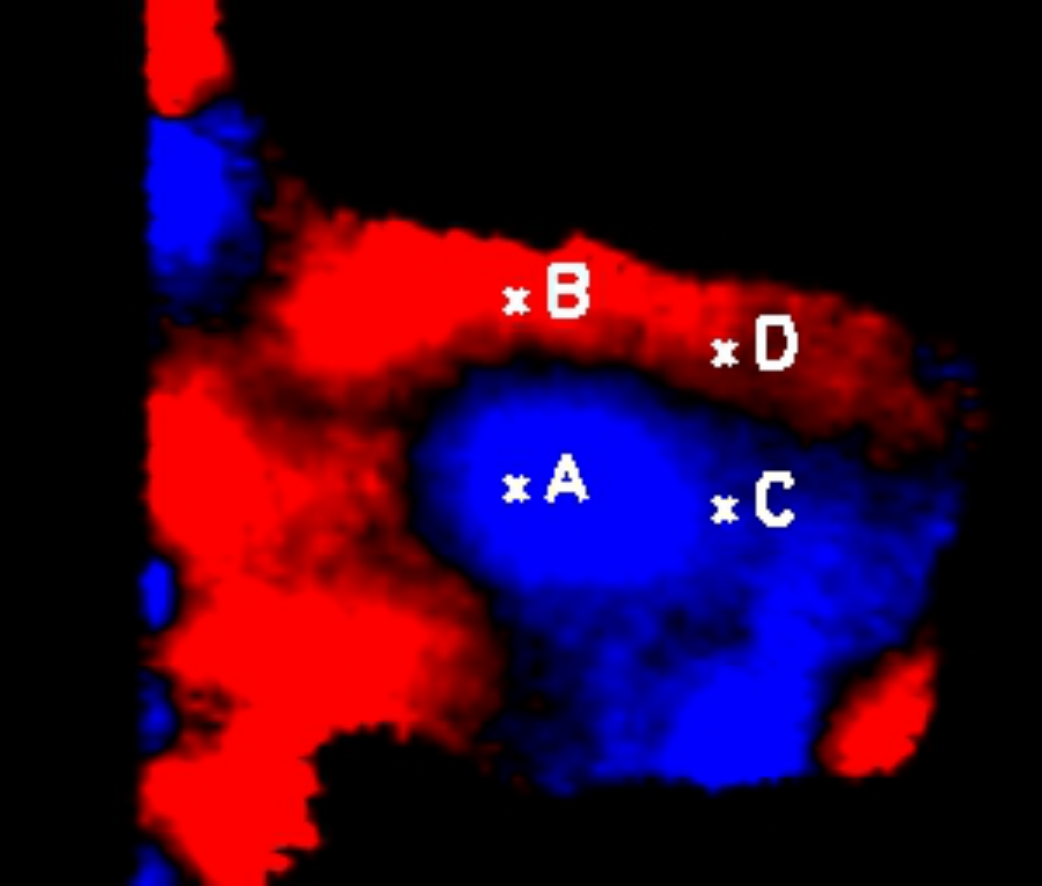}
     \includegraphics[width=7cm,angle=-90]{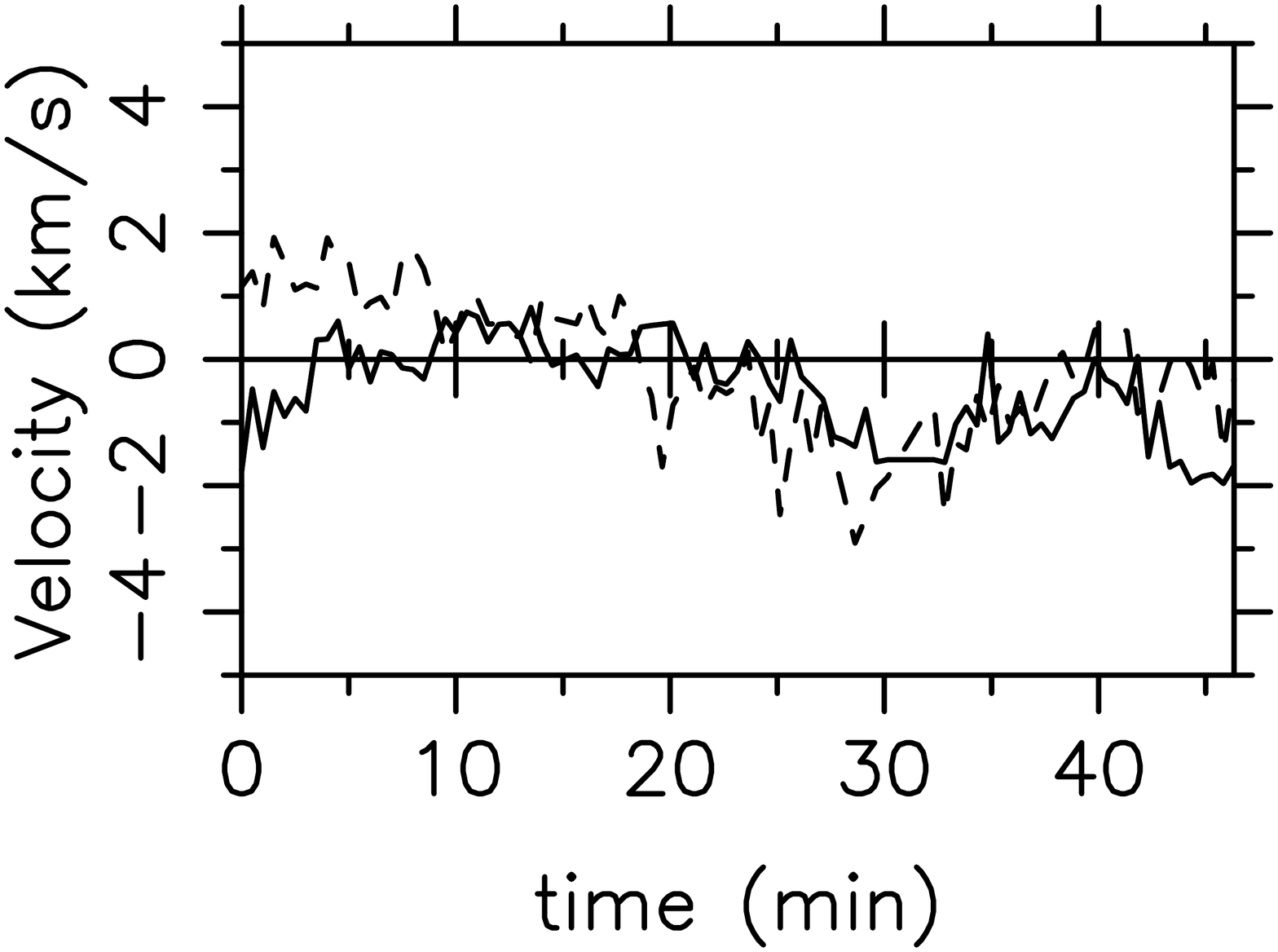}
      \includegraphics[width=7cm,angle=-90]{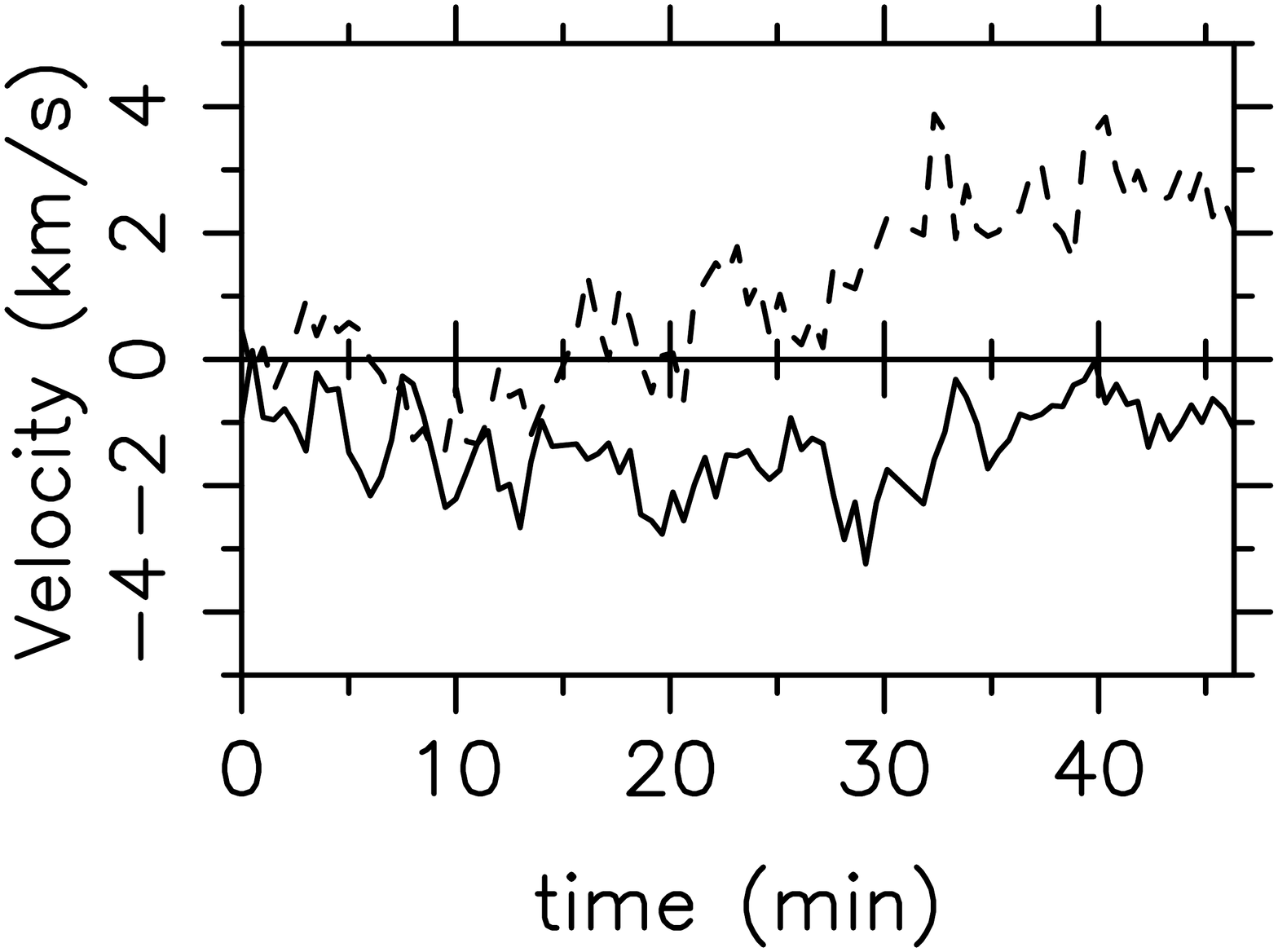}
      \caption{Evolution of Doppler shifts vs. time between  12:06  UT and 12:53 UT. \textit{Top panel:}  H$\alpha$ Doppler map, showing four points in the tornadoes (A, B, C and D), where the variation of the Doppler shift was measured. \textit{Middle panel:} Variation of the Doppler shift at points A (solid line) and B (dashed line). \textit{Bottom panel:} Variation of the Doppler shift at points C (solid line) and D (dashed line). See the colour bar in Figure \ref{Doppler1}.}
         \label{Doppler3}
   \end{figure}
   
%Figure 8
 \begin{figure}
   \centering
   \includegraphics[width=\hsize]{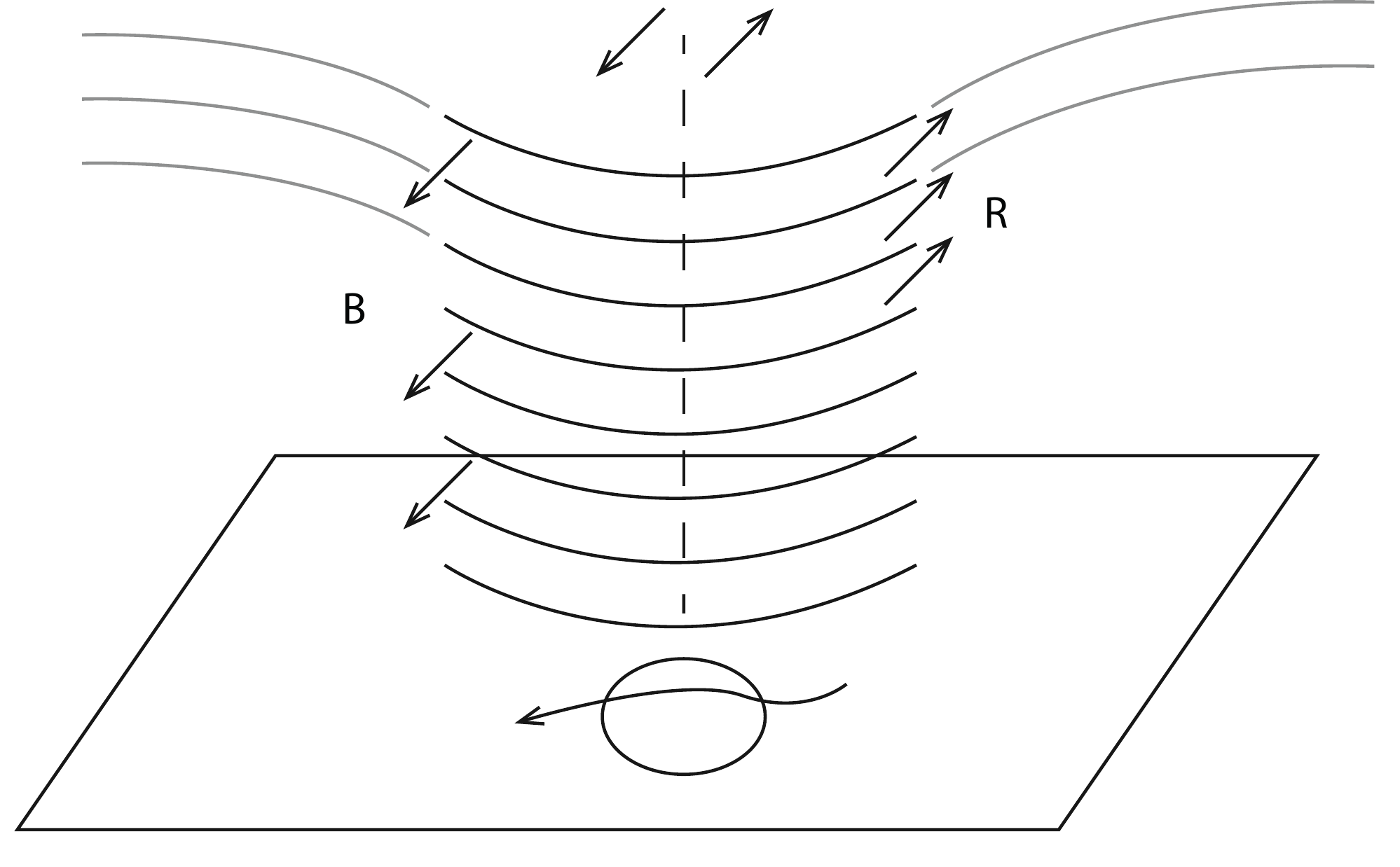} 
      %\caption{Sketch of a tornado model  for two times during the oscillations. The images show dipped magnetic field lines of a flux tube attracted by  a parasitic polarity (indicated here by the circle in the horizontal plan). `R' means redshift and `B' means blueshift, as seen along that line of sight. The arrows (straight and turning)  indicate the direction of motion of the magnetic field lines.}
      \caption{Sketch of a tornado model for a time during the oscillations. The image shows dipped magnetic field lines of a flux tube attracted by  a parasitic polarity (indicated here by the circle in the horizontal plane). `R' means redshift and `B' means blueshift, as seen along that line of sight. The arrows (straight and turning)  indicate the direction of motion of the magnetic field lines.}
         \label{cartoon}
   \end{figure}

\section{H$\alpha$ data}

The tornado is part of  a prominence   that was observed as a filament for a few days  in survey images from Meudon. The filament is oriented along a north-south meridian.  The filament was the target of a coordinated campaign as it was crossing the limb on September 24, 2013, with coordination between ground-based instruments and  the spectrograph aboard the space mission   IRIS.

Time sequences of H$\alpha$ MSDP observations obtained at the Meudon Solar Tower  have already been analysed in  \citet{Schmieder2014b} and \citet{Heinzel2015}.  These two papers are focused on the  part of the prominence  observed by  IRIS  (Figure \ref{MSDP}, white box). Here we consider  the tornado-like  structure located in the lower part  of the field of view (Figure \ref{MSDP}, red box).

\subsection{MSDP data processing}

The MSDP is a spectrograph with a wide slit acting as a field stop
\citep{Mein1991}.
In the Meudon MSDP, the size of the elementary field of view is 
465\arcsec $\times$ 60\arcsec. The pixel size is $0.6$\arcsec.
An optical system using prisms before the entrance window scans
the target by five successive images with small overlaps 
for adjustments by cross-correlations. The result is a 450\arcsec $\times$ 260\arcsec
full map recorded within 30 seconds. The size of the field
can be slightly different according to the spatial shifts 
deduced from correlations between elementary fields
and depending on seeing and coelostat motions.
The observations are obtained in an (x,y) non-heliographic coordinate system.
The top panel of Figure \ref{bissector} shows the map of prominence intensities at 12:22 UT.

%\begin{figure}[h]
%\begin{center}
%\resizebox{\hsize}{!}
%\includegraphics[angle=-90,width=8cm]{protu1222}
%\caption {MSDP Intensity map (x,y) of the prominence observed at 12/22 UT.}
%\end{center}
%\label{observ}
%\end{figure}

 In each of the nine channels of elementary frames
the wavelength is approximately constant along x, but varies along y.
The wavelength shift between two successive channels is 
 $\Delta\lambda=0.30$ $\AA$.
In channels 1 and 9, 
located practically outside the emission line, the $H\alpha$ prominence
is not visible anymore. 
The data processing includes geometrical calibrations to locate
the same solar points in all channels, and photometric calibrations
providing continuity between overlaps of successive channels.
Figure \ref{bissector} {\it bottom  panel} shows a typical prominence profile 
from channels 2 to 8.
It is defined by a cubic interpolation.
 Doppler shifts are determined by a bisector method
between points of equal intensity and wavelength distance, 

\begin{equation}
D_k = k  \Delta\lambda 
.\end{equation}

In this paper, observations are processed with $k = 2$.
Prominence pixels are selected by bisector intensities larger than
a given threshold (10\% of line centre disc intensity
near the limb).

\subsection{Typical velocity amplitudes}

The zero Doppler shift is determined by using a mean value across 
the full prominence visible in the 450\arcsec $\times$ 260\arcsec field of view, 
including the upper part of the prominence, north of the tornado (Figure \ref{bissector}, {\it top panel}).
In this way, we obtain accurate velocities relative to a structure 
that is much more extended than the tornado itself.

To estimate the relative accuracy of obtained Doppler velocities $v$,
we need an estimate of typical amplitudes. 
We take as an example the  time sequence starting at 12:06 UT and ending at 12:21 UT.
For each time, we can compute the root mean square of measured 
velocities over the full field of the prominence as follows:

\begin{equation}
\sigma (t)= (\langle V(x,y,t)^2\rangle_{x,y})^{1/2}
.\end{equation}

The $\sigma (t)$  values are averaged over the 30 maps of the time sequence 
to provide the typical amplitude
\begin{equation}
A =  \langle \sigma (t)\rangle_t = 2.16 $ $ $km/s$
\end{equation}

%\begin{equation}
% \Bigg \langle \left( \Big \langle v^2 \Big \rangle ^{1/2} \right) \Bigg \rangle_t = 2.16   \mathrm{km/s}
%\end{equation}

The accuracy of the result also depends on the zero level
of the Doppler shifts.
We consider the differences between the average velocities 
of successive maps  and their average over the same sequence.
The corresponding $rms$ is
\begin{equation}
 \delta v_0 = 0.11 $ $ $km/s$ 
.\end{equation}

It is less than 5\% of the typical velocity amplitude 
and can be neglected as a first approximation.  

\subsection{Comparison with Gaussian interpolation}

 It is interesting to check whether velocity measurements 
are dependent on the interpolation method. 
Figure \ref{bissector} shows the bisector method
applied to a tornado line profile 
with cubic spline interpolation (solid line).
We added a dashed line showing the Gaussian function determined by the
three points of the highest intensities (channels 4, 5, and 6). 
In this example, the agreement is very good in the core of the line  
near the points where the bisector method determines
the Doppler shift. 

To get more quantitative values, we compared departures
for all points of the time sequence starting at 12:06 UT.
Table 1 gives the  root mean squares of departures
in four strips parallel to the solar limb.
Each strip is 10\arcsec\ wide. 
The distances from solar limb to strip centres are 
increasing from 15\arcsec\ to 45\arcsec.

\begin{table}[h]
\caption{Root mean squares of departures between measured velocities in km s$^{-1}$
obtained with cubic and Gaussian interpolations for strips of 
different distances from the solar limb.}
\begin{center}
\begin{tabular}{|l|c|c|c|c|}
\hline
   & 15\arcsec & 25\arcsec & 35\arcsec & 45\arcsec\\ 
\hline
 $rms[V(x,y,t)-V_{Gauss}(x,y,t)]$    & 0.48  & 0.37 & 0.23 & 0.27 \\ 
\hline 
\end{tabular}

\end{center}
\end{table}

We see that the departures are globally decreasing with distance from the limb,
until values less than 0.3 km s$^{-1}$, that is only 14\% of
 typical velocity amplitudes.

Slightly higher values are observed close to the limb. They may be due 
to profiles, including several solar structures, or  velocity gradients
that cannot be represented by only one Gaussian profile.

\subsection{Upper estimate of data noise effects}

 It is possible to get upper estimates of data noise effects
by comparing measured velocities in neighbouring points
in time and space. We consider four points of the tornado 
(see Figure \ref{Doppler3} Sect. 3.7) during the full time sequence from 12:06 UT to 12:53 UT. 

\begin{table}[h]

\caption{Root mean squares of departures between measured velocities in km s$^{-1}$
at neighbouring times and positions for 4 tornado points A, B, C, and D (see Figure \ref{Doppler3}).}
\begin{center}

\begin {tabular}{|l|c|c|c|c|}
\hline
  &  A  &  B  &  C  &  D  \\
\hline
$rms[V(x,y,t+ dt)-V(x,y,t)]$ & 0.53  &  0.66  &  0.53  &  0.65  \\
$rms[V(x+dx,y,t)-V(x,y,t)]$ & 0.15  &  0.27  &  0.17  &  0.27  \\
$rms[V(x,y+ dy,t)-V(x,y,t)]$ & 0.13  &  0.27  &  0.18  &  0.26  \\
\hline
\end{tabular}

\end{center}
\end{table}

Table 2 shows the root mean squares of departures between
measured velocities in km s$^{-1}$ at successive times $t$ and $t+dt$ , 
and neighbouring points $x$ and $x+dx$ or $y$ and $y+dy$,
where $dt$ is almost always equal to 30 s
and $dx=dy$ equal to 0.5\arcsec.

Departures corresponding to displacements $dx$ and $dy$ are low,
and less than 0.3 km s$^{-1}$ or 14\% of typical amplitudes.
Departures corresponding to successive times are larger.
We see later (Figure \ref{Doppler3}, Sect. 3.7) that velocities versus time
are not really stochastic.
Evolving solar structures can account for such departures.

It is also possible to estimate noise effects directly with a CCD camera,
although it depends very much on the profile intensity from pixel to pixel.
In the case of Figure \ref{bissector}, the corresponding signal-to-noise ratio
at the intensity level of points used in the bisector method is around 60.
Since the slope of line profile is near 0.17 km s$^{-1}$ for 1\% relative
intensity fluctuations,
and since the Dopplershift is deduced from the half sum of 2 wavelengths,
we can estimate CCD noise effects at 0.20 km s$^{-1}$.
This value is slightly reduced, in fact, if we take
interpolations used in data reduction into account.

Finally, we can conclude that effects due to data noise are probably
smaller than 0.3 km s$^{-1}$.

%_______
       
\subsection{Scattering effects}

In a previous paper concerning prominences \citep{Gunar2012}, 
large parts of the solar limb
were visible outside the prominence, so that it was possible 
to observe scattered H$\alpha$ profiles  directly. 
This is not the case in the present set of data.
It can be noted that the far wings of the line, near the continuum, 
have  almost no tornado/prominence  contribution. 

We correct for scattered light in the following way:
we consider cuts crossing the limb in channels 1 and 9
along lines parallel to the longer edge of the field stop 
(the $x$ direction in Figure \ref{bissector}, {\it top panel}).
These cuts provide  intensity curves  that  approximately correspond
to constant wavelengths. 
These can be used as standard distributions of stray light 
outside the solar disc for any wavelength, 
on the condition that they are normalized by disc intensity and
measured at a given distance from the limb (for example 20\arcsec)
along the same direction of constant wavelength.

In all MSDP channels, a correction for stray light 
is obtained by subtracting such normalized intensity functions
along cuts parallel to the longer edge of field stop.  
This correction is a lower limit for the stray light level 
because the disc radius is larger in the H$\alpha$ profile 
than it is in the continuum.
It is also a rough approximation because the relative limb darkening 
is not exactly the same at all wavelengths.

\subsection{Relative intensity and profiles}
 
After having  applied all calibrations and corrections,
we did not use the coordinate system of observations (x,y) further.
For Figure \ref{MSDP},  where   the  tornado is presented 
in  a general context,  we have co-aligned the MSDP  and SDO 
images and we adopted  heliographic coordinates.  There is a rotation of -10 degrees with the (x,y) coordinates.
To obtain the evolution of the Doppler shifts versus time 
we computed time distance diagrams and again rotated the field of view 
of the data cubes to have one axis parallel to the limb. 
It is this reference system that is used in Figures \ref{profil},  \ref{Doppler1}, \ref{Doppler2}, and \ref{Doppler3}.

Figure \ref{profil} (top panel) presents an example of the observations 
of the tornado  with the MSDP, and Figure \ref{profil} (bottom  panel)
 shows some typical profiles of H$\alpha$  in the centre and 
edges of the tornado.
 The tornado  shows a double structure at its top, as indicated by red arrows
 in Figure \ref{profil} (top panel).

\subsection{Doppler shifts}

We analysed the spatial variation of H$\alpha$  Doppler shifts  
at $\pm 0.30 \AA$ in the tornado to see whether or not we can detect a signature of rotation in the cool plasma (10$^4$ K), as is concluded in some studies \citep{Orozco2012,Wedemeyer2013}.

Figures \ref{Doppler1} and \ref{Doppler2} show the dynamical behaviour of  the prominence
%5 cuts through the prominence velocity maps
between 12:06 UT and 12:53 UT and between 13:04 UT and 14:02 UT, respectively. The upper panels of these two figures show snapshots of the Doppler maps at different times, revealing the evolution of the redshift and blueshift patterns in the prominence.  The values of the Doppler shifts are coherent in areas (5 to 10\arcsec) that are much larger than the spatial resolution.
Such coherent patterns of  similar cell sizes have already been found in legs of prominences \citep{Schmieder2010}. 
 The lower panels show the evolution of these velocities over time for the five cuts indicated in the upper panels. 
%Blue cells become red and return to blue over a period of around one hour. 
These time slice Doppler shift  diagrams are notably different depending on which cut is being considered.
Looking at the temporal evolution,    it appears that the blue regions become red over quasi-periodic intervals, and vice versa.  After $\sim$ 20 minutes to 30 minutes  the  blue region has again become a red region.

%The time slice diagrams are quite different versus the cut that we used  crossing the tornado. 
If we consider only one cut during the first part of the sequence, i.e.  cut 1 as seen in Figure \ref{Doppler1}, we could conclude that we are seeing a rotation of the structure with redshifts on one side of the axis and blueshifts on the other, as found by \citet{Orozco2012}.  
% From the beginning  of first cut in Figure \ref{Doppler2}, we could deduce a rotation motion around the axis of the structure, as shown in Orozzco Su\'arez et al (2012).
 However, at the end of the sequence and in other cuts, the Doppler shift patterns are difficult to explain in terms of rotation. 
We can also see that the two parallel structures detected in the intensity map (Figure \ref{profil}) behave differently.

Figure \ref{Doppler3} shows the velocity behaviour versus time for four selected points (A, B, C, D) in these two structures.  Large period oscillations with periods of 40 to 60 minutes period are clearly seen.
The velocities versus time
are not really stochastic. The curves corresponding to points C and D,
 in particular, seem to exhibit some periods of the order of 4 or 6 minutes, 
which are much larger than the time interval between successive exposures.
Evolving solar structures can account for such oscillations.
%The general pattern seems to be better interpreted as slow oscillations with a period of approximatively one hour.

However the velocity measured at each pixel is  a mean value of a number of thread velocities along the line of sight. The H$\alpha$ line is optically thin and many structures are integrated along each line of sight, and these structures may have different velocities. Models of multiple threads with random velocity distribution  have correctly reproduced H$\alpha$ profiles \citep{Gunar2010,Gunar2012,Labrosse2016}.

\subsection{Electron and ion densities}
The top panel of Figure \ref{profil} shows the H$\alpha$ image of the tornado at 12:22 UT, and corresponds to the  prominence shown in the box in  Figure 1.  At the top  of the tornado we can distinguish two structures, shown in Figure \ref{profil} with two red arrows, which correspond to the  two structures visible in AIA  that appear to oscillate (Figure, \ref{time}, top cut).

Three characteristic  H$\alpha$ profiles are  shown in Figure \ref{profil} (bottom  panel). The units of intensity are  erg s$^{-1}$ cm$^{-2}$ sr$^{-1}$ Hz$^{-1}$. These profiles have been calibrated using the standard reference profiles defined by \citet{David1961}. %The profiles have been calibrated by using the standard David profile.  
 At disc centre the central intensity of the H$\alpha$ profile  is I$_c = 4.077 \times 10^{-5}$ erg s$^{-1}$ cm$^{-2}$ sr$^{-1}$ Hz$^{-1}$. The ratio of the peak intensity of the centre of the prominence, I$_{peak}$, to disc centre intensity, I$_c$, is I$_{peak}$/I$_c$ =  0.25/4.077 = 6.1\%\\
 According to the graph in \citet{Wiik1992} (their figure 10) this corresponds to an  electron density, $n_e$, in the range $2 - 4 \times 10^{10}$ cm$^{-3}$.
 The height at which these values are measured (black arrow, Figure \ref{profil}, lower panel) is at a much lower altitude than the 45 Mm that was used in the calculations of \citet{Wiik1992}, so these values may be reduced slightly.
% The height of the measures  (black arrow in Figure \ref{profil}) are not at such  an altitude as it is computed in Wiik et al 1992 (45000km), may it can reduce slightly  the values.
 
The integrated intensity  (I$_{int}$ =  $\sum$ I$_{\nu}$  $ d \nu$  = $\sum$ I$_{\nu}$  c /$\lambda ^2$ $d \lambda$)  of the tornado from H$\alpha$ profiles were computed for the following three locations in the tornado:\\
South  edge :  $1.1 \times 10^{5}$ erg s$^{-1}$ cm$^{-2}$ sr$^{-1}$\\
Centre : $1.5 \times 10^{5}$ erg s$^{-1}$ cm$^{-2}$ sr$^{-1}$\\
North edge : $0.74 \times 10^{5}$ erg s$^{-1}$ cm$^{-2}$ sr$^{-1}$\\

These values were compared to the model of \citet{Gouttebroze1993} to see what total hydrogen  density, $n_H$, it gives. The values are close to some of the models and seem to suggest that we have a total hydrogen  density of $n_H = 1.8 \times 10^{11}$ cm$^{-3}$ for $T$ = 6000 K, $p$ = 0.2 dyne cm$^{-2}$ and $D$ = 1000 km, which is similar to the values for $T$, $D$ and $p$ that were estimated in \citet{Levens2016a}. The peak intensity from the model is $3.01 \times 10^{-6}$ erg s$^{-1}$ cm$^{-2}$ sr$^{-1}$ Hz$^{-1}$ (integrated intensity, $I_{int} = 1.14 \times 10^{5}$ erg s$^{-1}$ cm$^{-2}$ sr$^{-1}$), where we find a value of $2.5 \times 10^{-6}$ erg s$^{-1}$ cm$^{-2}$ sr$^{-1}$ Hz$^{-1}$ at the centre of the tornado. This is consistent with previously measured values. It leads to a plasma-$\beta$ of 3 $\times$ 10$^{-3}$, confirming that the magnetic pressure dominates over the gas pressure. 
The  total hydrogen  density is an important parameter for the estimation of the Alfv\'{e}n speed in the prominence plasma in the context of waves that could be used to explain the oscillations.

\section{Discussion and conclusions}
A tornado-like structure was observed on September 24, 2013 by AIA  with a rotating period of  around 90 minutes as it crossed the limb.
 %A prominence was observed in multiple wavelengths on September 24, 2013. It appears that a part of  this prominence  has  an apparent rotation with a period of  around 90 minutes in the AIA 193 \AA\, movie.
 %  (Figures \ref{Doppler1}, \ref{Doppler2}, \ref{Doppler3}) \ref{time}).
 At the Meudon Solar Tower this tornado  prominence  was observed for a few hours with the MSDP 
 spectrograph in H$\alpha$  with a high cadence (two images per minute).
 Because of its concept of a large open slit at the entrance of the spectrograph, the MSDP allows us to have  3D data cubes of intensity in two coordinates and $\lambda$, and to recover Doppler shift data cubes in two coordinates and  time by 
processing a long sequence of observations.
% It is relatively unique data, and is very important for this kind of study i.e. measuring oscillations.

 We analysed the spatial variation of the Doppler
 shifts in the prominence to see if we can detect rotational  motion in the cool plasma.
The Doppler shift maps  present a pattern of small areas of alternating blueshifts and redshifts, which evolve over time.
 We can detect a double structure parallel to its central axis. Looking at the temporal 
 evolution,  it appears that the blue region becomes red  with a quasi-periodicity of 40 to 60 minutes.  Superimposed onto these oscillations are small amplitude oscillations with periods between  4 and  6 minutes.
 
 %ver a period of the order of  around one  hour. This pattern evolution can be  interpreted  as slow magnetosonic mode oscillations  and  not due to a rotation motion around the central axis.
 The behaviour of the H$\alpha$ Doppler shift  pattern  is very different from the EIS observations, 
 which show persistent blueshifts and redshifts symmetrically about the tornado axis \citep{Su2014,Levens2015}. 
 The capabilities of the MSDP instrument offer 
 the possibility of looking at the temporal evolution of the Doppler shifts at any position in the FOV. With EIS, the few studies that seem to point to 
rotation are much more limited in the sense that the sit-and-stare is only at one location. 
This would be equivalent to taking, for example the time-slice diagram at  cut 1 in Figure \ref{Doppler1}  or cut 3 in Figure  \ref{Doppler2} and saying that this is proof of rotation. Now, with the MSDP we show that, by looking at several locations, the data is not consistent with rotation. 
This is, so far, a unique analysis of Doppler shifts in a 3D data cube (x,y,t) with a high cadence (two images per minutes).

 The H$\alpha$ velocity behaviour gives some constraints on the models proposed to interpret the rotation of AIA tornadoes.
A first scenario could be to interpret the quasi-periodic pattern  observed by AIA as being due to kink oscillations such as in \citet{Poedts2015}.   Such a scenario leads those authors to derive  some  physical characteristics of their  tornado. In particular  they found a magnetic field  around B $ = 5$ gauss \citep{Poedts2015}.

%\citet{Poedts2015}  gave three different mechanisms to explain the observed oscillations in a tornado viewed by AIA.  The quasi-periodic displacement of the tornado axis could be due to MHD kink oscillations, the rotation of tornado threads around a common axis and the rise  of a twisted magnetic tornado. The solution given by \citet{Poedts2015} for MHD kink oscillations {leads to  a value of the magnetic field of $B = 5$ Gauss using the parameters of their tornado: length, period, and typical value for the electron density.}

%The wavelength of the fundamental harmonic is four times the actual length of the tornado, leading to a value 4 $\times$ 50 Mm = 200 Mm. The kink speed could be around $c_k$ = 70 km s$^{-1}$ inside the tornado. The Alfv\'{e}n is $v_a = c_k/\sqrt{2} \approx 50$ km s$^{-1}$, and assuming an electron density of $n_e = 4.7 \times 10^{10}$ cm$^{-3}$ gives a value of .

However we already have some indications about the magnetic field measured in a tornado by THEMIS \citep{Schmieder2015,Levens2016b}.  The magnetic field strength has been found to be between 10 and 45 gauss and is not as weak as in quiescent prominences (B = 5 gauss).  
%The orientation of the magnetic field is mainly horizontal, parallel to the solar limb.
Considering the results that we get in the present analysis of the tornado of September 24, 2013 we can estimate the phase speed of the apparent motions.
 The quasi-periodicity of the oscillations perpendicular to its vertical axis seen with AIA  is around 90 minutes.
Preliminary calculations of transverse oscillations for  cool plasma would lead to the following estimation: 
% (the  next paragraphs are the  computation of Ofman):
 the sound speed is $\sim$ 7 km s$^{-1}$ (based on $T = 6000$ K for the prominence plasma).
The Alfv\'{e}n speed is $\sim$ 127 km s$^{-1}$ (using  a mean value $B$ = 25 G, pressure $p$ = 0.3 dyn cm$^{-2}$, and $T$ = 6000 K, giving a total hydrogen density of $n_H = 1.8  \times 10^{11}$ cm$^{-3}$). If this is a kink oscillation, then the phase speed should be higher by $\sim \sqrt{2}$, assuming that the density outside the prominence is very small compared to the prominence material density, and we get the kink speed, $c_k$ = 179 km s$^{-1}$.

The phase speed of the wave from the fundamental mode can be estimated \citep[from e.g. coronal seismology, see review by][]{Nakariakov2005} as $c = 2L/P = 2 \times 10^{5}/(90 \times 60) = 37$ km s$^{-1}$, where $P$ is the oscillation period and $L$ is the length of the oscillating thread.  If the thread is not oscillating at the fundamental mode, then the wavelength could be shorter by the mode number factor. For example, assuming a full wavelength, we get $c$ = 18.5 km s$^{-1}$.  
 The estimated phase speed from the apparent oscillation is too high to  match  either the sound nor Alfv\'{e}n speed  according to  
   the  values of  temperature, density of the cool  H$\alpha$ plasma that we determined spectroscopically, and the estimated magnetic field strength  of tornadoes.   This solution could be valid only for extreme values of  B or L. The L value would be the full length of the prominence plus the tornado (see Figure \ref{MSDP}) and B should be  lowered by a factor 2 to 4.    We do not believe that B  in tornadoes can be reduced to   5 gauss because of our  B measurements in tornadoes presented in previous papers   \citep{Schmieder2015,Levens2016b}.   The conclusion of  this analysis  is  that the pure kink mode wave interpretation is difficult to be  justified in our case. 
 
 %However, many assumption and approximations have been made in the application of the seismology method, and the result could be significantly different with a more detailed analysis of the oscillations.
%{The expressions we have used for the estimate of the phase speed were derived for a slab, or cylinder, of oscillating plasma using linear theory - simplified coronal seismology. Clearly, in a prominence the magnetic geometry is much more complex than a slab or cylinder, and it is likely that there are no pure normal modes, but rather coupled oscillations in an inhomogeneous medium with additional effects of non-linearity. 
%This can result in different phase speed than predicted by simplified coronal seismology methods. 
%The determination of the modes and their couplings could potentially be improved with detailed 3D MHD modelling of the oscillations, given that we can include the real magnetic field and density structure in the model. However this is beyond the scope of the present paper. }

Another solution would be  to  consider that the oscillations are  from  plasma motions along the magnetic field. The orientation of the magnetic field in tornadoes was found mainly horizontal \citep{Schmieder2015,Levens2016b}.
If we   consider longitudinal waves along the field lines, as in \citet{Terradas2015} and \citet{Luna2016},  the Doppler shifts  would indicate a component of the velocity along the line of sight,  which is more or less perpendicular to the direction of the horizontal field, if  we assume that the tornado is connected to the northern feet of the prominence by horizontal field lines, parallel to the solar limb. This would mean that the velocity of the plasma along the field is much larger than the 2 km s$^{-1}$ that has been measured. The pendulum model, where plasma is moving along rigid magnetic field lines, could explain the oscillations \citep{Luna2016}. This orientation could account for both the plane-of-sky motion seen in AIA and the oscillation signatures seen in the Doppler maps from MSDP.
 The period of the oscillation would depend on the curvature of the dipped field lines that support the plasma. The different periods registered along the prominence could be due to the different curvatures.  In that case the oscillations observed by AIA and differences in the direction of the Doppler velocities in H$\alpha$ across the tornado would be caused by counter-streaming, as was discussed in \citet{Panasenco2014}. It is, however, difficult to justify that the field lines are rigid and do not move in such a dynamic atmosphere that is governed by the magnetic pressure. 
 
 %Therefore a combination of these effects could be causing the motions that are seen here.

We now discuss  MHD sketch  models.
The transverse displacement  shown in AIA 
should   also be along the line of sight, and it should deform the magnetic field slightly. The azimuth would be an interesting parameter to study over time. However we have no evolution of the magnetic field parameters over time during the present observation because of the low temporal resolution of THEMIS; it takes one hour to measure the Stokes parameters for a field of view of 120\arcsec $\times$ 160\arcsec.  To combine the observations of the  transverse oscillations seen in AIA and the oscillations of the H$\alpha$ plasma, we need to consider horizontal magnetic field lines that have an angle with the line of sight, somewhere around 45$^\circ$. The dip field lines  would therefore not be in the plane of the sky, but viewed from an angle.% , if we consider there is counter-streaming in the prominence. In this model, the plasma would simply be a pendulum, oscillating in the dipped magnetic field \citep{Luna2015}.

 Tornadoes are probably  the legs of prominences, as has been suggested previously \citep{Wedemeyer2013,Levens2016a}, and could be modelled as a pile up of dips supporting the cool plasma in the corona. The legs (barbs, footpoints) are directly related to parasitic polarities \citep{Aulanier1998}. \citet{Schmieder2014b} demonstrate that the polarities related to the anchorage of the prominence legs are   at the borders of the supergranules and generally at the convergence point between two supergranules. The diffusion of polarities appearing in the internetwork 
%is  around 10 hour (30Mm/2.x 0.2 km/s) (30Mm/3600*0.2x2)= 15/ 1.8). C
or cancelling flux  are  permanent motions in the dynamic photosphere.
% Oscillation  behaviour would be induced by photospheric motions that would be  transformed to magneto-acoustic  waves as a response of the local eigenfrequency.
The piles of dips would move from left to right with some kind of oscillatory  motions (Figure \ref{cartoon}).

 The main conclusion of this paper is that the H$\alpha$  Doppler shift pattern  of  the tornado observed 
%characterized by in H$\alpha$ using a large slit spectrograph  (MSDP) observed 
simultaneously by AIA  in 193 \AA\  cannot be interpreted in terms of rotation.
 %as well as in terms of pure, linear MHD mode waves. 
 We therefore look forward to new observing campaigns involving ground-based instruments and  \textit{Hinode} (SOT and EIS) and IRIS to measure the Doppler shifts over a large temperature range to better understand the dynamical coupling between different plasmas in and around the prominence. The MHD simulations of oscillations in a realistic, three-dimensional magnetic field, and density model of a prominence would be suitable to disentangle the different mechanisms that have been suggested by these observations.

\begin{acknowledgements}
The authors would like to thank  the anonymous referee for his/her fruitful comments, R. Lecocguen, D. Crussaire and the team at the MSDP for acquiring the H$\alpha$ observations. {We thank G. Aulanier for fruitful discussions and the sketch of a `tornado', which has been drawn by Sylvain Cnudde. We deeply thank L. Fletcher for giving many comments on the manuscript to improve its clarity. } P.J.L. acknowledges support from an STFC Research Studentship ST/K502005/1. N.L. acknowledges support from STFC grant ST/L000741/1. L.O. would like to acknowledge support by NASA Cooperative Agreement grant NNG11PL10A to CUA. The AIA data are provided courtesy of NASA/\textit{SDO} and the AIA science team.
\end{acknowledgements}

%-------------------------------------------------------------------

%\bibliographystyle{aa}
%\bibliography{bibliography}

\begin{thebibliography}{44}
\expandafter\ifx\csname natexlab\endcsname\relax\def\natexlab#1{#1}\fi

\bibitem[{{Antolin} {et~al.}(2015){Antolin}, {Okamoto}, {De Pontieu},
  {Uitenbroek}, {Van Doorsselaere}, \& {Yokoyama}}]{Antolin2015}
{Antolin}, P., {Okamoto}, T.~J., {De Pontieu}, B., {et~al.} 2015, \apj, 809, 72

\bibitem[{{Anzer} \& {Heinzel}(2005)}]{Anzer2005}
{Anzer}, U. \& {Heinzel}, P. 2005, \apj, 622, 714

\bibitem[{{Aulanier} \& {Demoulin}(1998)}]{Aulanier1998}
{Aulanier}, G. \& {Demoulin}, P. 1998, \aap, 329, 1125

\bibitem[{{Berger}(2014)}]{Berger2014}
{Berger}, T. 2014, in IAU Symposium, Vol. 300, IAU Symposium, ed.
  B.~{Schmieder}, J.-M. {Malherbe}, \& S.~T. {Wu}, 15--29

\bibitem[{{Chae}(2010)}]{Chae2010}
{Chae}, J. 2010, \apj, 714, 618

\bibitem[{{David}(1961)}]{David1961}
{David}, K.-H. 1961, \zap, 53, 37

\bibitem[{{Dud{\'{\i}}k} {et~al.}(2008){Dud{\'{\i}}k}, {Aulanier}, {Schmieder},
  {Bommier}, \& {Roudier}}]{Dudik2008}
{Dud{\'{\i}}k}, J., {Aulanier}, G., {Schmieder}, B., {Bommier}, V., \&
  {Roudier}, T. 2008, \solphys, 248, 29

\bibitem[{{Dud{\'{\i}}k} {et~al.}(2012){Dud{\'{\i}}k}, {Aulanier}, {Schmieder},
  {Zapi{\'o}r}, \& {Heinzel}}]{Dudik2012}
{Dud{\'{\i}}k}, J., {Aulanier}, G., {Schmieder}, B., {Zapi{\'o}r}, M., \&
  {Heinzel}, P. 2012, \apj, 761, 9

\bibitem[{{Engvold}(2008)}]{Engvold2008}
{Engvold}, O. 2008, in IAU Symposium, Vol. 247, Waves: Oscillations in the
  Solar Atmosphere: Heating and Magneto-Seismology, ed. R.~{Erd{\'e}lyi} \&
  C.~A. {Mendoza-Briceno}, 152--157

\bibitem[{{Gouttebroze} {et~al.}(1993){Gouttebroze}, {Heinzel}, \&
  {Vial}}]{Gouttebroze1993}
{Gouttebroze}, P., {Heinzel}, P., \& {Vial}, J.~C. 1993, \aaps, 99, 513

\bibitem[{{Gun{\'a}r} {et~al.}(2012){Gun{\'a}r}, {Mein}, {Schmieder},
  {Heinzel}, \& {Mein}}]{Gunar2012}
{Gun{\'a}r}, S., {Mein}, P., {Schmieder}, B., {Heinzel}, P., \& {Mein}, N.
  2012, \aap, 543, A93

\bibitem[{{Gun{\'a}r} {et~al.}(2010){Gun{\'a}r}, {Schwartz}, {Schmieder},
  {Heinzel}, \& {Anzer}}]{Gunar2010}
{Gun{\'a}r}, S., {Schwartz}, P., {Schmieder}, B., {Heinzel}, P., \& {Anzer}, U.
  2010, \aap, 514, A43

\bibitem[{{Heinzel} \& {Anzer}(1999)}]{Heinzel1999}
{Heinzel}, P. \& {Anzer}, U. 1999, \solphys, 184, 103

\bibitem[{{Heinzel} {et~al.}(2015){Heinzel}, {Schmieder}, {Mein}, \&
  {Gun{\'a}r}}]{Heinzel2015}
{Heinzel}, P., {Schmieder}, B., {Mein}, N., \& {Gun{\'a}r}, S. 2015, \apjl,
  800, L13

\bibitem[{{Kosugi} {et~al.}(2007){Kosugi}, {Matsuzaki}, {Sakao}, {Shimizu},
  {Sone}, {Tachikawa}, {Hashimoto}, {Minesugi}, {Ohnishi}, {Yamada}, {Tsuneta},
  {Hara}, {Ichimoto}, {Suematsu}, {Shimojo}, {Watanabe}, {Shimada}, {Davis},
  {Hill}, {Owens}, {Title}, {Culhane}, {Harra}, {Doschek}, \&
  {Golub}}]{Kosugi2007}
{Kosugi}, T., {Matsuzaki}, K., {Sakao}, T., {et~al.} 2007, \solphys, 243, 3

\bibitem[{{Labrosse} \& {Rodger}(2016)}]{Labrosse2016}
{Labrosse}, N. \& {Rodger}, A.~S. 2016, \aap, 587, A113

\bibitem[{{Lemen} {et~al.}(2012){Lemen}, {Title}, {Akin}, {Boerner}, {Chou},
  {Drake}, {Duncan}, {Edwards}, {Friedlaender}, {Heyman}, {Hurlburt}, {Katz},
  {Kushner}, {Levay}, {Lindgren}, {Mathur}, {McFeaters}, {Mitchell}, {Rehse},
  {Schrijver}, {Springer}, {Stern}, {Tarbell}, {Wuelser}, {Wolfson}, {Yanari},
  {Bookbinder}, {Cheimets}, {Caldwell}, {Deluca}, {Gates}, {Golub}, {Park},
  {Podgorski}, {Bush}, {Scherrer}, {Gummin}, {Smith}, {Auker}, {Jerram},
  {Pool}, {Soufli}, {Windt}, {Beardsley}, {Clapp}, {Lang}, \&
  {Waltham}}]{Lemen2012}
{Lemen}, J.~R., {Title}, A.~M., {Akin}, D.~J., {et~al.} 2012, \solphys, 275, 17

\bibitem[{{Levens} {et~al.}(2015){Levens}, {Labrosse}, {Fletcher}, \&
  {Schmieder}}]{Levens2015}
{Levens}, P.~J., {Labrosse}, N., {Fletcher}, L., \& {Schmieder}, B. 2015, \aap,
  582, A27

\bibitem[{{Levens} {et~al.}(2016{\natexlab{a}}){Levens}, {Schmieder},
  {Labrosse}, \& {L{\'o}pez Ariste}}]{Levens2016a}
{Levens}, P.~J., {Schmieder}, B., {Labrosse}, N., \& {L{\'o}pez Ariste}, A.
  2016{\natexlab{a}}, \apj, 818, 31

\bibitem[{{Levens} {et~al.}(2016{\natexlab{b}}){Levens}, {Schmieder},
  {L{\'o}pez Ariste}, {Labrosse}, {Dalmasse}, \& {Gelly}}]{Levens2016b}
{Levens}, P.~J., {Schmieder}, B., {L{\'o}pez Ariste}, A., {et~al.}
  2016{\natexlab{b}}, \apj, 826, 164

\bibitem[{{L{\'o}pez Ariste}(2015)}]{Lopez2015}
{L{\'o}pez Ariste}, A. 2015, in IAU Symposium, Vol. 305, IAU Symposium, ed.
  K.~N. {Nagendra}, S.~{Bagnulo}, R.~{Centeno}, \& M.~{Jes{\'u}s
  Mart{\'{\i}}nez Gonz{\'a}lez}, 207--215

\bibitem[{{Luna} {et~al.}(2015){Luna}, {Moreno-Insertis}, \&
  {Priest}}]{Luna2015}
{Luna}, M., {Moreno-Insertis}, F., \& {Priest}, E. 2015, \apjl, 808, L23

\bibitem[{{Luna} {et~al.}(2016){Luna}, {Terradas}, {Khomenko}, {Collados}, \&
  {de Vicente}}]{Luna2016}
{Luna}, M., {Terradas}, J., {Khomenko}, E., {Collados}, M., \& {de Vicente}, A.
  2016, \apj, 817, 157

\bibitem[{{Mackay} {et~al.}(2010){Mackay}, {Karpen}, {Ballester}, {Schmieder},
  \& {Aulanier}}]{Mackay2010}
{Mackay}, D.~H., {Karpen}, J.~T., {Ballester}, J.~L., {Schmieder}, B., \&
  {Aulanier}, G. 2010, \ssr, 151, 333

\bibitem[{{Mein}(1991)}]{Mein1991}
{Mein}, P. 1991, \aap, 248, 669

\bibitem[{{Mghebrishvili} {et~al.}(2015){Mghebrishvili}, {Zaqarashvili},
  {Kukhianidze}, {Ramishvili}, {Shergelashvili}, {Veronig}, \&
  {Poedts}}]{Poedts2015}
{Mghebrishvili}, I., {Zaqarashvili}, T.~V., {Kukhianidze}, V., {et~al.} 2015,
  \apj, 810, 89

\bibitem[{{Nakariakov} \& {Verwichte}(2005)}]{Nakariakov2005}
{Nakariakov}, V.~M. \& {Verwichte}, E. 2005, Living Reviews in Solar Physics, 2

\bibitem[{{Ofman} {et~al.}(2015){Ofman}, {Knizhnik}, {Kucera}, \&
  {Schmieder}}]{Ofman2015}
{Ofman}, L., {Knizhnik}, K., {Kucera}, T., \& {Schmieder}, B. 2015, \apj, 813,
  124

\bibitem[{{Ofman} {et~al.}(1998){Ofman}, {Kucera}, {Mouradian}, \&
  {Poland}}]{Ofman1998}
{Ofman}, L., {Kucera}, T.~A., {Mouradian}, Z., \& {Poland}, A.~I. 1998,
  \solphys, 183, 97

\bibitem[{{Okamoto} {et~al.}(2015){Okamoto}, {Antolin}, {De Pontieu},
  {Uitenbroek}, {Van Doorsselaere}, \& {Yokoyama}}]{Okamoto2015}
{Okamoto}, T.~J., {Antolin}, P., {De Pontieu}, B., {et~al.} 2015, \apj, 809, 71

\bibitem[{{Okamoto} {et~al.}(2007){Okamoto}, {Tsuneta}, {Berger}, {Ichimoto},
  {Katsukawa}, {Lites}, {Nagata}, {Shibata}, {Shimizu}, {Shine}, {Suematsu},
  {Tarbell}, \& {Title}}]{Okamoto2007}
{Okamoto}, T.~J., {Tsuneta}, S., {Berger}, T.~E., {et~al.} 2007, Science, 318,
  1577

\bibitem[{{Orozco Su{\'a}rez} {et~al.}(2012){Orozco Su{\'a}rez}, {Asensio
  Ramos}, \& {Trujillo Bueno}}]{Orozco2012}
{Orozco Su{\'a}rez}, D., {Asensio Ramos}, A., \& {Trujillo Bueno}, J. 2012,
  \apjl, 761, L25

\bibitem[{{Panasenco} {et~al.}(2014){Panasenco}, {Martin}, \&
  {Velli}}]{Panasenco2014}
{Panasenco}, O., {Martin}, S.~F., \& {Velli}, M. 2014, \solphys, 289, 603

\bibitem[{{Schmieder} {et~al.}(2010){Schmieder}, {Chandra}, {Berlicki}, \&
  {Mein}}]{Schmieder2010}
{Schmieder}, B., {Chandra}, R., {Berlicki}, A., \& {Mein}, P. 2010, \aap, 514,
  A68

\bibitem[{{Schmieder} {et~al.}(2013){Schmieder}, {Kucera}, {Knizhnik}, {Luna},
  {Lopez-Ariste}, \& {Toot}}]{Schmieder2013}
{Schmieder}, B., {Kucera}, T.~A., {Knizhnik}, K., {et~al.} 2013, \apj, 777, 108

\bibitem[{{Schmieder} {et~al.}(2004){Schmieder}, {Lin}, {Heinzel}, \&
  {Schwartz}}]{Schmieder2004}
{Schmieder}, B., {Lin}, Y., {Heinzel}, P., \& {Schwartz}, P. 2004, \solphys,
  221, 297

\bibitem[{{Schmieder} {et~al.}(2015){Schmieder}, {L{\'o}pez Ariste}, {Levens},
  {Labrosse}, \& {Dalmasse}}]{Schmieder2015}
{Schmieder}, B., {L{\'o}pez Ariste}, A., {Levens}, P., {Labrosse}, N., \&
  {Dalmasse}, K. 2015, in IAU Symposium, Vol. 305, IAU Symposium, ed. K.~N.
  {Nagendra}, S.~{Bagnulo}, R.~{Centeno}, \& M.~{Jes{\'u}s Mart{\'{\i}}nez
  Gonz{\'a}lez}, 275--281

\bibitem[{{Schmieder} {et~al.}(2014{\natexlab{a}}){Schmieder}, {Roudier},
  {Mein}, {Mein}, {Malherbe}, \& {Chandra}}]{Schmieder2014b}
{Schmieder}, B., {Roudier}, T., {Mein}, N., {et~al.} 2014{\natexlab{a}}, \aap,
  564, A104

\bibitem[{{Schmieder} {et~al.}(2014{\natexlab{b}}){Schmieder}, {Tian},
  {Kucera}, {L{\'o}pez Ariste}, {Mein}, {Mein}, {Dalmasse}, \&
  {Golub}}]{Schmieder2014}
{Schmieder}, B., {Tian}, H., {Kucera}, T., {et~al.} 2014{\natexlab{b}}, \aap,
  569, A85

\bibitem[{{Su} {et~al.}(2014){Su}, {G{\"o}m{\"o}ry}, {Veronig}, {Temmer},
  {Wang}, {Vanninathan}, {Gan}, \& {Li}}]{Su2014}
{Su}, Y., {G{\"o}m{\"o}ry}, P., {Veronig}, A., {et~al.} 2014, \apjl, 785, L2

\bibitem[{{Terradas} {et~al.}(2015){Terradas}, {Soler}, {Luna}, {Oliver}, \&
  {Ballester}}]{Terradas2015}
{Terradas}, J., {Soler}, R., {Luna}, M., {Oliver}, R., \& {Ballester}, J.~L.
  2015, \apj, 799, 94

\bibitem[{{Wedemeyer} {et~al.}(2013){Wedemeyer}, {Scullion}, {Rouppe van der
  Voort}, {Bosnjak}, \& {Antolin}}]{Wedemeyer2013}
{Wedemeyer}, S., {Scullion}, E., {Rouppe van der Voort}, L., {Bosnjak}, A., \&
  {Antolin}, P. 2013, \apj, 774, 123

\bibitem[{{Wiik} {et~al.}(1992){Wiik}, {Heinzel}, \& {Schmieder}}]{Wiik1992}
{Wiik}, J.~E., {Heinzel}, P., \& {Schmieder}, B. 1992, \aap, 260, 419

\bibitem[{{Xia} {et~al.}(2014){Xia}, {Keppens}, {Antolin}, \&
  {Porth}}]{Xia2014}
{Xia}, C., {Keppens}, R., {Antolin}, P., \& {Porth}, O. 2014, \apjl, 792, L38

\end{thebibliography}

\end{document}